\documentclass[flushrt,preprint]{aastex}
\input{psfig.tex}  

\begin{document}

\slugcomment{Astronomical Journal, in press}

\shorttitle{The distance to the LMC from the RR Lyrae stars}
\shortauthors{Clementini et al.}

\title{DISTANCE TO THE LARGE MAGELLANIC CLOUD: 
THE RR LYRAE STARS
\footnote{Based on observations collected at the
European Southern Observatory, proposal numbers 62.N-0802, 66.A-0485,
and 68.D-0466} }

\author{Gisella Clementini$^2$, Raffaele Gratton$^3$, Angela Bragaglia$^2$,
Eugenio Carretta$^3$, Luca Di Fabrizio$^{4}$, and Marcella Maio$^2$}
\author{}
\affil{$^2$INAF - Osservatorio Astronomico di Bologna, Via Ranzani 1, 
  40127 Bologna, ITALY}
\affil{$^3$ INAF - Osservatorio Astronomico di Padova, Vicolo dell'Osservatorio
  5, 35122 Padova, ITALY}
\affil{$^4$ INAF - Centro Galileo Galilei \& Telescopio Nazionale Galileo, 
PO Box 565, 
38700 S.Cruz de La Palma, Spain}
\authoremail{gisella@bo.astro.it, gratton@pd.astro.it, angela@bo.astro.it,
carretta@pd.astro.it, difabrizio@tng.iac.es, s\_maio@astbo4.bo.astro.it}

\begin{abstract}
New photometry and spectroscopy for more than a hundred RR Lyrae stars in  two
fields  located close to the bar of the Large Magellanic Cloud (LMC) are used
to derive new accurate  estimates of the  average magnitude, the local
reddening, the luminosity-metallicity  relation, and of the distance to the
LMC. The average apparent luminosity of the RR Lyrae's with complete  $V$ and
$B$ light  curves  is 
$<V(RR)>=19.412 \pm 0.019$ ($\sigma$=0.153),
$<B(RR)>=19.807 \pm 0.022$ ($\sigma$=0.172) in our field A (62 stars), and 
$<V(RR)>=19.320 \pm 0.023$ ($\sigma$=0.159),  
$<B(RR)>=19.680 \pm 0.024$ ($\sigma$=0.163) in our field B (46 stars).
The average $V$ apparent luminosity of the clump stars in the same areas is
0.108 and 0.029 mag  brighter than the RR Lyrae level               
($<V_{clump}>=19.304 \pm 0.002$ and 19.291 $\pm 0.003$, in  field A: 6728
stars, and B: 3851 stars, respectively). 
Metallicities  from low resolution spectra obtained with the Very Large
Telescope have been derived for 101 RR Lyrae stars, finding an average value of
[Fe/H]=$-1.48 \pm 0.03$ ($\sigma$=0.29, on Harris 1996 metallicity scale).
An estimate of the reddening within the two fields was obtained (i) from 
Sturch (1966) method applied  to the fundamental mode pulsators (RRab's) with
known metal abundance;  and (ii) from  the colors of the edges of the
instability strip defined by the full sample of RR Lyrae  variable stars. We
obtained : $E(B-V)$=0.116$\pm$0.017, and  0.086$\pm$0.017 mag  in field A and
B,  respectively, with a clearcut indication of a 0.03 mag differential 
reddening betweeen the two fields.  We find that reddening in field A is 0.028
mag  smaller  than  derived by OGLE-II in the same area.
On average, the new reddenings are also 0.035 mag larger than derived from
Cepheids with  projected distances within 2 degrees from the centers of our
fields.
The new metallicities were combined with   the apparent average V$_0$
luminosities to determine the slope of the  luminosity-metallicity  relation
for the RR Lyrae stars. We derived  $\Delta  {\rm M_V(RR)}/\Delta$[Fe/H]=
0.214$\pm$0.047,  with no clear evidence for the change in   slope at
[Fe/H]=$-1.5$, recently suggested by evolutionary/pulsation and  Horizontal
Branch models. 

The dereddened apparent average luminosity of the  RR Lyrae's defined by the
present photometry is $<V(RR)>_0=19.064 \pm 0.064$ at [Fe/H]=$-1.5$.

When coupled with the absolute magnitude derived from  the Baade-Wesselink
and   the statistical parallaxes methods (M$_{V}(RR)$=0.68$\pm$0.15  and
0.76$\pm$0.13 mag  at  [Fe/H]=$-$1.5), both methods  known to favour the {\it
short} distance scale,  this value leads to distance moduli for the  LMC of
$\mu_{{\rm LMC}}$=18.38 $\pm$ 0.16 and  $\mu_{{\rm LMC}}$=18.30 $\pm$ 0.14,
respectively. If we use instead the absolute magnitude from the new Main 
Sequence Fitting  of Galactic globular clusters (M$_{V}(RR)$=0.61$\pm$0.07 mag 
at  [Fe/H]=$-$1.5; Gratton et al. 2002a) we derive $\mu_{{\rm LMC}}$=18.45
$\pm$ 0.09.

The average $I$ apparent luminosity of the clump stars derived by the  present
photometry is  $<I_{clump}>=18.319 \pm 0.002$ and 18.307 $\pm 0.003$, in  field
A ($\sigma$=0.190, 6728 stars) and B ($\sigma$=0.184, 3851 stars), 
respectively. These values once  corrected for our new reddening  estimates
lead to $<I>_0$=18.12 $\pm$0.06 mag  and move the clump  distance modulus to
the LMC to 18.42$\pm 0.07$ and 18.45 $\pm$0.07,   when Udalski [2000a, ApJ,
531, L25] or Popowski [2000, ApJ 528, L9]   metallicity-$I$  luminosity
relations for the clump stars are adopted.

All these values are only 1 $\sigma$ shorter  than provided by the Population I
distance indicators, and allow to  {\bf reconcile the short and  long distance
scale} on a common value for the  distance modulus of the LMC of $\mu_{{\rm
LMC}}$=18.515$\pm$0.085 mag. 

\end{abstract}

\keywords{distance scale -- Magellanic Clouds -- stars: oscillations -- 
stars: variables: other (RR Lyrae) -- techniques: photometry -- 
techniques: spectroscopy} 

\newpage

\section{Introduction}

The Large Magellanic Cloud (LMC) is widely considered as a benchmark in the 
definition of the  astronomical distance scale; however no general consensus
has been reached  so far on its actual distance and  a dichotomy at a 0.2-0.3
mag level seems  to persist between {\it short} and {\it long} distances to the
LMC as derived  from  Population I and II indicators.
Population I distance indicators have a preference to  cluster around a {\it	
long} distance modulus for the LMC in the range from 18.5 to 18.7 mag. 		
An internal dichotomy is presented instead by the Population II  indicators	
with most indicators based on "field stars"  (Baade-Wesselink,	
B-W,  and Statistical Parallax methods, in particular) yielding a  {\it short}	
distance modulus of 18.3, while indicators based on "cluster stars" (e.g.	
the Main Sequence Fitting method: Gratton et al. 1997,  Carretta et	
al. 2000b, Gratton et al. 2002a; the globular clusters  dynamical models;
or the pulsational properties of RR Lyrae's in globular clusters: Sandage 1993a)
support a {\it  longer} modulus in the range from 18.4 to 
18.6.		
Indeed, the well-known discrepancy between {\it short} and   {\it long} 	
distance scales derived from old, Population II stars is still an  unsolved
issue (see e.g. Gratton et al. 1997), in spite of the  impressive improvements	
in measuring distances achieved thanks to the  Hipparcos mission.  The average	
difference between the two scales (0.25-0.30 mag) translates into a difference	
of about 3-4$\times 10^9$ years in the corresponding age of the Galactic	
globular clusters and in turn in the age of the Universe.			

\subsection{The RR Lyrae distance scale to the LMC}

The distance to the LMC from Population II objects is finally founded  on the
luminosity of the RR Lyrae variables: observations provide the "apparent
luminosity" of the LMC RR Lyrae's:  m(RR$_{\rm LMC}$). Various techniques
provide the absolute  luminosity of the RR Lyrae's: M(RR). Distance to the LMC
simply follows  from the well known relation between apparent and absolute
magnitudes. The first step actually includes two issues: (a) the derivation/use
of a relation between the absolute luminosity of RR Lyrae stars and
metallicity; and (b) the derivation of the apparent magnitude of RR Lyrae stars
at a given metal abundance. Both issues are addressed in this paper with the
help of new, photometric and spectroscopic data we have obtained for more than
100 RR Lyrae stars in the bar of the LMC.

Various estimates of the  {\bf apparent luminosity} of  RR Lyrae stars in the
LMC existed so far, with small but significant differences.
(i) Walker (1992a) published average luminosities in the Johnson system  for RR
Lyrae's in  7 clusters of the LMC; the mean dereddened apparent magnitude of
these stars is 18.95 $\pm$ 0.04 mag (182 variables) at an  average metal
abundance of [Fe/H]= $-1.9$, and for an average reddening value of $<{\rm
E(B-V)}>$=0.09 mag. A slightly different value of  18.98 $\pm$ 0.04 mag (160
variables) and  $<{\rm E(B-V)}>$=0.07 mag is obtained if NGC 1841, a cluster
suspected  to be about 0.2 mag closer to us (Walker  1992a), is discarded.
(ii) The MACHO microlensing experiment has led to the  discovery of about 8,000
RR Lyrae's in the bar of the LMC  (Alcock et al. 1996, hereinafter referred to 
as A96), among which 73 double-mode pulsators (Alcock et  al. 1997, hereinafter
referred to as A97). The average dereddened  apparent magnitude of the RR
Lyraes   published by A96 for a subsample of 500  of these variables is 19.09
at  [Fe/H]=$-$1.6 (the most frequent metallicity value in A96 spectroscopic 
abundances of 15 RR Lyrae's in their sample) and for an assumed reddening value
of $E(B-V)$=0.1. Recently, MACHO photometry has been re-calibrated to the
standard Johnson-Cousins  photometric system (Alcock et al. 1999, hereinafter
referred to as  A99) superseding all prior calibrations. Alcock et al. (2000,
hereinafter referred to as A00)  present the color-magnitude diagram of 9
million stars (9M CMD) contained in  22 MACHO fields covering the LMC bar, and
provide  also median  luminosity and metallicity of about 680 RRab's: 
$V$=19.45 mag, and [Fe/H]=$-1.6$, respectively, and the revised  average 
luminosity of A97 RRd's: $<V> =19.327\pm 0.021$, based on the new photometric
calibration.
(iii) A further estimate of the apparent dereddened average  luminosity of  RR
Lyrae's in the LMC comes from observations   obtained during the second phase
of the Optical Gravitational Lensing  Experiment, OGLE-II (Udalski, Kubiak \& 
Szyma\'nski 1997). Udalski (1998a) quotes an apparent dereddened average
luminosity of  $<$V$_0$(RR)$> $=18.86 $\pm$0.04 mag at an average metallicity
of [Fe/H]=$-1.6\pm 0.2$ (see Table 6 of  Udalski 1998a) and for  assumed
reddening values in the range from $E(B-V)$=0.17 to 0.20 mag  (see Table 2 of  
Udalski et al. 1998) based on 104  RR Lyrae's observed in OGLE-II fields 
LMC\_SC14, 15, 19 and 20. This value was lately revised to $<$V$_0$(RR)$>
$=18.94 $\pm$0.04 mag (Udalski et al. 1999b), on the claim that extinction was
slightly  overestimated in Udalski (1998a). Extrapolating back from this
revised  $<$V$_0$(RR)$>$, the new adopted E(B--V) shoud be  about 0.025 mag
smaller than in Udalski (1998a). Udalski (2000b) has furtherly revised  the
OGLE-II apparent dereddened average luminosity of the RR Lyrae in the LMC bar
to $<$V$_0$(RR)$>$=18.91 $\pm$0.01 mag,  apparently based  on the total sample
of Lyrae variables detected  by OGLE-II in the LMC  bar, at [Fe/H]=$-$1.6 and
for  E(B$-$V)=0.143 mag. These new $<$V$_0$(RR)$>$ and reddening are
inconsistent  with both Udalski (1998a) and Udalski et al. (1999b) values for
these quantities; we further note that  if the 0.3 dex difference in
metallicity between Udalski (2000b) and Walker (1992a) samples is taken into
account,  Walker's RR Lyraes should be 0.06 mag brighter and not 0.04 (or 0.07)
mag  fainter than Udalski (2000b). Both authors claim that their photometric
zero-points are good to 0.02-0.03 mag, so  this could suggest that Walker's
reddening is too small or Udalski's too large by E(B$-$V)=0.03--0.04 mag.

Estimates of the {\bf absolute magnitude} of the RR Lyrae stars come from 
the B-W method  (Liu \& Janes 1990; Jones et al. 1992; Cacciari, Clementini 
\&  Fernley 1992; Skillen et al. 1993; Fernley 1994), and from the 
Hipparcos  based
Statistical parallaxes (Fernley et al. 1998a;   Tsujimoto et al. 1998;  Gould
\& Popowski 1998), applied to field RR Lyrae's.

Fernley et al. (1998b) provide a summary of previous B-W  analyses and discuss
the slope and zero-point of the RR Lyrae  absolute magnitude -- metallicity
relation. They  quote a zero-point of M$_V$(RR)=0.98$\pm$0.15 mag (at
[Fe/H]=0.0),  and a mild slope of 0.18$\pm$0.03 mag, given by the weighted
average of the B-W slope (0.20$\pm$0.04) and of results for 8 globular 
clusters (GCs) in M31 observed with the Hubble Space Telescope (Fusi Pecci et 
al. 1996: 0.13$\pm$ 0.07).  However, analysis of an enlarged
sample of 19 M31 GCs (Rich  et al. 2001) now suggests a steeper slope of
0.22 mag/dex. An even steeper slope of  0.30 mag/dex was found by   Sandage
(1993a)  from the pulsational properties of RR Lyrae stars.  Finally, a shallow
slope of $\sim$ 0.18-0.22 mag/dex has been obtained from  evolutionary HB
models (e.g. Caloi et al. 1997, Cassisi et al. 1998).  Moreover, recent
evolutionary/pulsation and HB models also suggest that the  slope  may be not
unique in the metallicity range spanned by the Galactic GCs  (Castellani et al.
1991, Cassisi et al. 1998, Caputo 1997, Demarque et al.  2000) with two
different regimes for metallicities below and  above [Fe/H]=$-1.5$.

Statistical parallaxes lead to 
M$_{\rm V}$(RR)=0.77$\pm$0.15 mag at [Fe/H]=$-$1.53 (Fernley et al. 1998a),
0.69 $\pm$0.10 mag at [Fe/H]=$-$1.58 (Tsujimoto et al. 1998), and
0.77$\pm$0.13 at [Fe/H]=$-$1.60 (Gould \& Popowski 1998), but do not 
provide information on the slope of the M$_{\rm V}$(RR)-[Fe/H] relation.

In the remaining part of this Section we assume a value of $\Delta  
{\rm M_V(RR)}/\Delta$[Fe/H]=0.2
mag/dex\footnote{We will return on the value of the slope of the 
luminosity-metallicity relation in Sect. 5.2.} (which is supported by both 
the B-W and the new M31 results) to
transform the  above M$_{\rm V}$(RR) values to the average metallicity of  the
LMC RR Lyrae stars in our sample. This is  assumed to be [Fe/H]=$-1.5 \pm 0.2$, 
which is   the most commonly 
adopted average value for the RR Lyrae variables in the LMC, and it was
confirmed by the  $\Delta$S metal abundances derived by Bragaglia et
al. (2001) for 6 of the  LMC RRd's in our sample.  This metallicity is also
consistent with A96, and with the average metal abundance we derived 
from low resolution spectroscopy of about 80\% of the 
RR Lyrae stars considered in this paper (see Sections 3.3 and 
Gratton et al. 2002b). 

We thus derive M$_{\rm V}$(RR)=0.72 $\pm$ 0.15 mag as  average of the  B-W
(0.68 mag at [Fe/H]=
$-$1.5)\footnote{The B-W determination of the absolute
luminosity of the Galactic field  RR Lyrae's is being revised in order to test
the effects on this technique of the most recent model atmospheres  (Kurucz
1995; Castelli 1999; Canuto \& Mazzitelli 1992) with  various approximations in
the treatment of convection, different values of  turbulent velocity and more
complete and accurate opacity tables, as well as  the use of instantaneous
gravity along the pulsation cycle. 
Preliminary  results (Cacciari et al. 2000) give  M$_{\rm V}$(RR)=0.56 mag, 
for one star at [Fe/H]=$-$1.5 (RR Cet); this is about 0.12 mag brighter than
found from  previous studies (e.g. Fernley et al. 1998b). A similar analysis is
presently  being performed on a larger sample  of field RR Lyrae's at
[Fe/H]=$-$1.5 (Cacciari, Clementini \&  Castelli 2003, in preparation), and on
RR Lyrae stars in the globular cluster M3.} and of the statistical parallaxes
results (0.76 at [Fe/H]=$-$1.5). This value, when combined with  Walker's
apparent luminosity for the LMC cluster RR Lyrae's leads to a  distance modulus
for the LMC of $\mu_{{\rm LMC}}$=18.32 (at [Fe/H]=$-$1.5 dex) in better
agreement with the {\bf short} distance scale. On the other hand, MACHO's
revised value ($V$=19.45 mag at [Fe/H]=$-$1.6)  leads to  a {\bf longer}
modulus of 18.44 mag for the LMC, but   further complicates the scenario since
it also seems to suggest that an  intrinsic difference of $\sim 0.1$ mag might
exist between field and cluster RR Lyrae's. In fact, if allowance is given for
the  0.3 dex difference in metallicity between Walker's and A96/A00 samples 
and according to $\Delta {\rm M_V(RR)}/\Delta$[Fe/H]=0.2,   A00 apparent
luminosity for the field RR Lyrae's is 0.12 mag {\bf fainter} than Walker's
apparent luminosity for the cluster variables\footnote{Even with the steep  
slope of $\Delta {\rm M_V(RR)}/\Delta$[Fe/H]$\sim$0.3, supported by Sandage
(1993a) there is  still a $-$0.10 mag difference between Walker's and A00, and
a +0.13 mag difference  between Walker's and Udalski (2000b),  reddening
corrected, apparent  luminosities for the LMC RR Lyrae's.}.
Finally, if the bright average magnitude proposed by Udalski (2000b)
is correct ($<$V$_0$(RR)$>$=18.91 $\pm$0.01 mag at [Fe/H]=$-$1.6), 
the distance modulus to the LMC would be as short as 
$\mu_{{\rm LMC}}$=18.21. This shows that simply considering different
recent determinations for the average magnitude of the RR Lyrae stars,
the range in the distance modulus values for the LMC is larger than 0.2 mag.
This situation is clearly unsatisfactory and requires further 
inquiry of this issue.

\subsection{What may cause these differences in $<$V$_0$(RR)$>$?}

An intrinsic difference at $\sim 0.1-0.2$ mag level between the luminosity  of
HB stars in globular clusters and in the field was tentatively suggested by 
Gratton (1998) calibration of the absolute magnitude of field  horizontal
branch (HB) metal-poor stars  with good parallaxes from Hipparcos. Sweigart
(1997) evolutionary  models, which include some extra-mixing of He and other
heavy elements (C,N,O) might give some support to this hypothesis, since 
there is
observational evidence for extra-mixing in cluster red giants but not among
field stars (Gratton et al. 2000c). However, Gratton's suggestion was tested
qualitatively by Catelan (1998) and, lately and in a more quantitative way, by
Carretta,  Gratton \& Clementini (2000a) using the pulsational properties of a
selected  sample of field and cluster RR Lyrae's. Their results show that field
and  cluster variables, in our Galaxy, cannot be distinguished in the  $\Delta
\log {\rm P_{field-cluster}}$ -- [Fe/H] plane, so that the actual  existence of
a luminosity difference between field and cluster variables  seems very
unlikely.

Alternative explanations for the differences in the
apparent luminosity of the LMC RR Lyrae's found by Walker (1992a), A00, and 
Udalski (2000b) should be considered. We notice that:

(i) MACHO photometry is in non standard blue and red filters, which  are very
different from the Johnson passbands of Walker's cluster RR Lyrae
photometry (see fig. 1 of A99). 
A99 find that there is a systematic shift of $\Delta V \sim -0.035$ mag between
MACHO and other literature data with MACHO 
photometry being brighter (see fig.7 of A99). Additional concern also
arises from the accuracy  of the MACHO  photometry at the limiting magnitude
typical for the LMC RR Lyrae's (19-20 mag); the problem is clearly shown in
fig. 6 of A99. 
Comparison with MACHO is presented in Section 2.2, and the problem is 
considered at lenght in 
Di Fabrizio et al. (2002, hereinafter 
referred to as DF02), where we find that 
MACHO photometric zero-points, while apparently correct in some of their
fields, appear to present a mismatch in others (see Section 2.2).

Udalski et al. (1997) provide a detailed description of the instrumental set up
of OGLE-II. They briefly mention the OGLE-II filter passbands and comment that,
except for the ultraviolet filter, they are very close to the standard 
Johnson-Cousins UBVRI photometric system. Udalski et al. (2000) 
compared OGLE-II  apparent magnitudes and colors for a few
local standard stars in the field of NGC 1835  they have in commom with  Walker
(1993). They found a mean difference of $\Delta$V=--0.026$\pm$0.025 mag 
(based on
6 stars, Walker's magnitudes  beeing slightly  brighter), and
$\Delta$(B--V)=--0.026$\pm$0.028 (based on 4 star, Walker's colors being
bluer), but also noted that  for stars redder than (B--V)$\sim$1.2 mag OGLE
colors are redder by more than  0.1 mag. Udalski et al. (2000)
 attribute this discrepancy to the Walker 
B-band filter, that matches rather poorly the standard B-band (the color 
coefficient  of
the transformation is about 1.2, see Walker 1992b). 
However,  given the
small number of objects in common between Walker (1993) and Udalski et al.
(2000), any firm conclusion on the comparison between the two  photometries
seems premature, particularly at red colors. 
This point is considered more extensively in DF02, where we conclude 
that OGLE-II systematically overestimates apparent luminosities 
for stars close to the limit of their $B$, $V$ photometry, including RR Lyrae 
and clump stars.
 
(ii) It may be possible that due to projection effects combined to small 
number statistics, the globular clusters in Walker (1992a) sample are,  on
average, closer to us than the bar of the LMC. Indeed,  Walker (1992a) assumed
that the clusters were on the same plane as younger  populations, based on
similarity  in kinematics (Schommer et al. 1992). However, this assumption may
be  questioned in view of the large scatter of the individual objects in the
sky  (see Figure~1), and Walker himself cautions that at least one of the 
clusters in his sample  (NGC 1841) could actually be about 10\% closer to 
us.

(iii) A major crucial point is the actual reddening  towards and inside the
LMC. Walker (1992a; see his Table 1) derives reddening values from  a number of
methods  in each  cluster of his sample, including two techniques based on  the
pulsational properties of RR Lyrae's, namely the color of the edges of the 
instability strip, and Sturch's (1966) method.  These reddenings range from 
$E(B-V)$= 0.03 for the Reticulum cluster to $E(B-V)$= 0.18 for NGC 1841, with 
an average value of $E(B-V)$= 0.09 mag. However, in some clusters the adoption
of such reddening values results in a too blue dereddened instability strip  
(NGC 1835), suggesting that in these cases the reddening may be  overestimated.
In other clusters (NGC 1786, NGC 1841) the strip  may be not totally filled:
reddening  estimates are rather uncertain in these cases. A96 do not derive an
independent reddening estimate, but  simply  assume $E(B-V)$=0.1 mag, based on
Bessel (1991)  re-examination of the reddening towards and inside the two
Clouds.

Udalski (1998a) adopts an average reddening value of $E(B-V)$=0.185 mag for his
RR Lyrae's  sample, but Udalski et al. (1999b) claim that this value is too
high.  They do not give the revised value,  but extrapolating back from the
revised  $<V_0(RR)>$, the new adopted $E(B-V)$ shoud be  about 0.16 mag.
Udalski et al. (1999a)  published reddening values for all the 21 LMC fields
observed by OGLE-II. They range from  E(B--V)=0.105 to 0.201 mag with an
average value of 0.143$\pm$0.017 mag. This average value is 0.042 and 0.017 mag
smaller than  the reddenings adopted for the LMC RR Lyrae stars  in Udalski
(1998a) and  Udalski et al. (1999b) respectively.

Finally, we recall that the reddening value derived from Cepheids in the bar 
of the LMC is E(B$-$V)= 0.07 mag (Laney \& Stobie 1994), i.e. $\sim$ 0.02-0.03 
mag smaller than  Walker's and A96, and about 0.07 mag smaller than Udalski
(2000b). This would be enough to justify most of the spread 
observed in different distance scales.
The work presented in this paper represents a step toward the solution of these
inconsistencies.

\section{Our new data-sets}

We have undertaken a photometric and spectroscopic observational campaign 
of two fields close to the bar of the LMC using the 1.5 m Danish, the 3.6 m, 
and the VLT ESO telescopes.
Within our program we have obtained  
(1) accurate $B, V$ and $I$ photometric light curves  directly
tied to the Johnson-Cousins photometric system for 152 variable stars 
in these areas, among which 125 RR Lyrae stars, and 
(2) low resolution spectra for 101 of the RR Lyrae stars,
among which 9 of A97 LMC  double-mode pulsators, 
and for more than 300 clump stars.
The main purposes of  the observing program were 
(a) the definition of the average luminosity, and 
(b) the measure of the metal abundance of RR Lyrae and clump stars in the  LMC, 
in order to investigate  the luminosity-metallicity and mass-metallicity
(for double-mode pulsators) 
relations of the  RR Lyrae's, and the metallicity distribution of the clump
stars in the LMC bar. The photometric data are presented in 
DF02.

Metal abundances for a first sample of 6 double-mode RR Lyrae stars  were
derived  using the $\Delta$S method (Preston 1959), and masses of the stars 
were estimated from new pulsational models and the Petersen diagram (Petersen
1973). The mass-metallicity distribution of the RR Lyrae's in  the LMC was
found to be indistinguishable from that obtained from Galactic globular
clusters, thus ruling out any intrinsic difference in mass which might induce
intrinsic differences between field and cluster RR Lyrae's. The results from 
this first spectroscopic study are described in Bragaglia et al. (2001). New
spectroscopic metal abundances for a sample comprising 80\% of the RR Lyrae
stars discussed in the present paper were obtained with FORS at the VLT in
December 2001. The metallicity derivation technique is  briefly outlined in
Section 3.3 and described in detail in Gratton et al.  (2002b). 

The present paper is devoted to the presentation of the results obtained from
the combination of the  photometric and specroscopic study of the RR Lyrae
stars. The new  photometry has allowed us to define a very precise average 
apparent luminosity for the RR Lyrae stars in the bar of the LMC, which can  be
compared  to Walker (1992a) estimate for the LMC cluster RR Lyrae's, as well as
to the mean apparent luminosity of the LMC clump stars found in the  same
fields. These data also allow an independent estimate of the reddening in these
regions of the LMC bar, with fundamental fall-backs on the  derivation of the
distance to the LMC. The individual metallicities are combined with   the
apparent V$_0$ luminosities to firmly establish the slope of the 
luminosity-metallicity  relation for RR Lyrae stars. 

The photometric and spectroscopic datasets used in this paper are briefly
described in Sections 2.1, 2.2, and 2.3 where we also provide a summary of the
results from the comparison of our, MACHO and OGLE-II photometries, based on
large samples of stars in common between the  three independent data-bases. The
properties of  the RR Lyrae variables in our sample (apparent average
luminosity, period and metallicity  distributions) are discussed in  Section
3.   Section 4 addresses the  problem of measuring the reddening in the LMC,
and   new estimates of $E(B-V)$ from the color of the edges of the instability
strip  defined by our RR Lyrae sample, and from Sturch method applied to the
RRab's with known metal abundances are presented.  In Section 5 we summarize
our results on the RR Lyrae average luminosity and on the LMC reddening, and
present  the luminosity-metallicity relation defined by  the RR Lyrae stars in
our sample, based on our new  photometric and spectroscopic analyses.
Comparison with Walker (1992a), MACHO and OGLE-II average luminosity of  the
LMC RR Lyrae stars is  presented in Section 6, where we also  briefly review
the use of the clump as distance indicator, and  compare  the average
luminosity of the RR Lyrae stars to the luminosity distribution  of the clump
stars in the same areas. Finally, in Section 7 we summarize our results and 
discuss the impact of the present new photometry, spectroscopy, and reddening
values on the derivation  of the  distance to the LMC. 

Photometric data, variable star identification and  period search procedures,
and the detailed comparison between our, MACHO, and OGLE-II photometries, are
fully described in  DF02, where we publish  the photometric catalogue with
multicolor light curves and the individual measurements for all the variable
stars in our  sample. Spectroscopic data, along with a detailed description of
the  metallicity derivation technique, and the individual spectroscopic
metallicities derived for the RR Lyrae stars are presented in Gratton et al.
(2002b), where we also provide details on the derivation of the RR Lyrae
luminosity-metallicity relation. Discussion of the luminosity and metallicity
distribution of clump stars, based on the spectroscopic database obtained with
the VLT, is postponed to a following paper in preparation.   

\subsection{The photometric dataset}

The photometric dataset used in this paper arises from observations taken with
the 1.54 m Danish telescope (La Silla, Chile)  in two separate observing runs
in January 1999 (four nights) and January 2001 (two nights), respectively. We
observed at two different positions, hereinafter  called field A and B,  close
to the bar of the LMC and contained in fields \#6 and \#13 of the MACHO
microlensing experiment. Field A is also about 40\% overlapped with OGLE-II
field LMC\_SC21 (Udalsky et al. 2000). In the 1999 run, observations were done
in the Johnson-Bessel $B$  and $V$ filters (ESO 450,  and 451, see ESO web
page), and we obtained 58 $V$ and 27 $B$ frames for field A, and 55 $V$ and 24
$B$ frames for field B.  In the 2001 run  observations were done in the
Johnson-Bessel-Gunn $B$, $V$ and $i$ filters  (ESO 450,  451, and 425)  and we
obtained 14 $V$, 14 $B$ and 14 $i$ frames for field A, and 15 $V$, 14 $B$, and
14 $i$ frames for field B.   Both nights of the 2001 run were photometric, and
with good  seeing conditions, while only one of the nights in 1999 was  
photometric. 

Reduction and analysis of the 1999  photometric data were done using the
package DoPHOT (Schechter, Mateo \& Saha 1993). Photometric reductions of the
2001 data were done using DAOPHOT/ALLSTAR II (Stetson 1996) and ALLFRAME
(Stetson 1994). A detailed  description of the reduction procedures can be
found in DF02. 

In the 2001 run, nights were superior to the ones in the 1999 run, and  a much
larger number of standard stars were observed. Therefore our entire photometric
dataset was tied to the standard  Johnson-Cousins photometric system through 
the absolute photometric calibration of the  2001 run. Calibration equations
are provided in DF02. They are    based on 127 measurements in two separate
nights of 27 standard stars selected from Landolt (1992) - Stetson (2000)
standard fields PG0918+029, PG0231+051, PG1047+003, and SA98,  with magnitude
and colors in the ranges   $12.773 < V < 17.729$, $-0.273 < B-V < 1.936$,
$-0.304< V-I < 2.142$. Photometric zero-points accuracies are of 0.02  mag in 
$V$ and 0.03 mag in $B$ and $I$,  respectively.

Variable stars were identified on the 1999 time-series dataset using the
program VARFIND, written for this purpose by P. Montegriffo.   Candidate
variable objects picked up interactively from  the scatter diagram produced by
VARFIND   were checked for variability using the program GRATIS  (GRaphycal
Analyzer TIme Series) a private software developed at the Bologna  Observatory
by P. Montegriffo, G. Clementini and L. Di Fabrizio. Confirmed variable stars
were then counteridentified on the 2001 frames using private software by P.
Montegriffo (see DF02 for  further details.)

We detected  152 variable objects in the two fields, and an additional 7
candidate variable objects of unknown type.   125 of the certain variables are
of RR Lyrae type  (115 single-mode and 10 double-mode RR Lyrae's, one of which
not previously  known from A97). Of course, the high fraction of RR Lyrae stars
among the identified variable stars is due to the type of observations  and
variable stars search we made. 

Periods for all the identified variable stars were defined using the program 
GRATIS on the instrumental differential photometry with respect to stable, well
isolated objects used as reference stars. Thanks to the two years base line of
our observations we were able to: (i) derive periods and epochs for all the 152
variables  accurate to better than the fourth- fifth decimal place; the
accuracy depends on the light curve data  sampling, which in the best cases is
72 $V$, 41 $B$, 14 $I$ data points for  variables  in field A, and 70 $V$, 38
$B$, 14 $I$ data points for variables in field B; (ii) identify the Blazhko
modulation of the light curve (Blazhko 1907) in about 15 \% of our RR Lyrae
stars; and  finally (iii)  to derive complete and well sampled $B$ and $V$
light curves for about 95\% of the RR Lyrae stars. Typical
accuracies\footnote{Accuracies are the average residuals from the Fourier best
fitting models of the light curves.} for the single-mode, non Blazhko variables
are 0.02-0.03 mag in $V$, and  0.03-0.04 mag in $B$.

Apart from RR Lyrae's, we also fully covered  the light curve of 4 Anomalous
Cepheids,  9 eclipsing binaries with short orbital  period (P$<$1.$^{\rm
d}4$),  6 classical Cepheids, and reasonably well sampled the light curves   of
the other 5 Cepheids. 

Examples of the light curves of an {\it ab}, {\it c} and {\it d} type RR 
Lyrae, as well as  of an anomalous Cepheid, a classical Cepheid and an
eclipsing binary in our  sample are shown in   Figure~2 and 3.  The full
catalogue of individual photometric measurements and light curves is provided
in DF02.

The location of the variable stars on the HR diagram of field A and B is shown
in Figure~4 and ~5, respectively, where they are plotted according to  their
intensity-average magnitudes and colors, and with different symbols 
corresponding to the various types. 

\subsection{Comparison with MACHO and OGLE-II photometries}

Our fields are contained in MACHO's fields \#6 and \#13, and there is about
40\% overlap between field A and OGLE-II field LMC\_SC21. This occurrence has
allowed us to make a detailed comparison between our,  MACHO, and OGLE-II
photometries, based on large samples of  stars, and without any assumption
about reddening. 

The comparison  with MACHO was restricted only to the variable stars in common,
since calibrated photometry of individual stars was readily available only for
the  variable star subsample. We found that we have  about 30\% and 20\% more
short period variables ($P < 4$ days) than  MACHO in field A and B,
respectively.   We also found that MACHO classification of some of the variable
stars in common did not match our classification (see details in DF02). 

The average $V$ difference, in the sense present photometry minus MACHO,  is
$-0.174$ mag ($\sigma=0.092$, 58 stars) in field A,  and $-0.010$ mag
($\sigma=0.085$, 40 stars) in field A. The  quite large systematic shift found
for variable stars in field A,  with MACHO luminosities being {\bf fainter}
than ours, is  rather surprising  and of no obvious explanation, since the
different treatment of the  background in the reduction package used by MACHO
(SoDoPhot)  is expected to produce {\bf brighter} magnitudes  than
DAOPHOT+ALLFRAME photometry (A99). Moreover, we notice  (i) that a shift by
+0.174 mag of field A photometry would result in an average  luminosity of the
RR Lyrae stars in this area about  0.24 mag fainter than A00 median luminosity
of the RRab's in the LMC,  and (ii) given the good agreement  existing in field
B,  it would imply a  rather unlikely difference of about 0.27 mag in the
average luminosity  of RR Lyrae stars in the two fields.

Our field A is 42.13 \% overlapped with OGLE-II field LMC\_SC21. In this common
area we measured 21,524 objects (in all three photometric bands),   our
limiting magnitude reaches about  1.5 mag fainter, and we resolved about 39,
26, and 22\% more stars  (in $B$, $V$ and $I$, respectively)  than OGLE-II.

There are  about 13,000 objects in common between the two photometries, with 
$V$ magnitude in the range from 16 to about 22.5 mag. Among these objects
OGLE-II reports 114 variable stars; 96 of them have a  counterpart in  our
database, but only 14 are identified as variable stars in our photometry. On
the other hand, in the same area we have 22 further variable stars all  having
a counterpart, but not  recognized as variable objects  by OGLE-II. A more
significant comparison with OGLE-II can be  achieved by restricting the 
sample  to the stars in common that  have photometric errors  smaller than 0.1
mag in all photometric passbands  in both databases (about  5400, 6500, and
7200 stars in $V, B$, and $I$, respectively).  Average residuals between the 
two photometries were computed  dividing objects in magnitudes bins; they are
provided in Table~5 of DF02. At the magnitude level of RR Lyrae and clump stars
($V\sim 19.4, B\sim 19.8, I \sim 18.8$; and
$V\sim 19.3, B\sim 20.2, I \sim 18.3$, respectively)
offsets are: $\Delta V$ = 0.06 ($\sigma_{V}$=0.03), 
$\Delta B$ = 0.03 ($\sigma_{B}$=0.04), $\Delta I$ = 0.07 ($\sigma_{I}$=0.05),
and $\Delta V$ = 0.06 ($\sigma_{V}$=0.03), 
$\Delta B$ = 0.03-0.04 ($\sigma_{B}$=0.04-0.05), 
$\Delta I$ = 0.07 ($\sigma_{I}$=0.04-0.05), with our photometry being
systematically {\bf fainter} than OGLE-II DoPhot photometry.
This is not unexpected since DoPhot is reported to give systematically 
brighter magnitudes for faint stars in crowded regions than DAOPHOT,  and since
we resolve many more  faint stars than OGLE-II  in the area in common.

The interested reader is referred to DF02 for a thorough discussion of the 
comparison with MACHO and OGLE-II photometries.

\subsection{The spectroscopic dataset}

Low resolution spectra (R$\sim$450) at minimum light for 6 of the double mode 
pulsators in our sample  were obtained in January 1999, with the Faint Object
Spectrograph  and Camera of the 3.6 m ESO telescope. We used the  classical
$\Delta S$ method (Preston 1959) to measure [Fe/H] abundances, as fully
described in Bragaglia et al. (2001).

A more extended spectroscopic survey  was conducted in 2001 with FORS at the
VLT. We obtained  low resolution spectra (R$\sim$815) for 80\% of the RR Lyrae
stars  discussed in this paper and for about 300 clump stars. Metal abundances
for the RR Lyrae stars were obtained using a revised version  of the  $\Delta
S$ technique: we estimated metallicities for individual stars by comparing 
directly the strength of  H lines and the K Ca II line with analogous data for
RR Lyrae stars in 3 Galactic Globular clusters of known metal abundance (namely
NGC 1851, M 68 and NGC 3201). An advantage of this  technique is that we do not
need an explicit phase correction; even if accuracy of our [Fe/H]
determinations is still a function of phase (as represented by  the strength of
the H lines): determinations accurate to about 0.2 dex are obtained only from
observations  taken far from maximum light. Note however that the allowed 
phase range is about two times larger than that required for a safe 
application of the original $\Delta$S method.  A detailed description of the
metallicity derivation technique can be found in Gratton et al. (2002b). 

\section{The RR Lyrae stars}
\subsection{The apparent luminosity}

The average apparent luminosities of Lyrae's with full coverage of the 
$V$ and $B$ light curves are:
$<V>=19.412\pm 0.019$ ($\sigma$=0.153, 62 stars), $<B>=
19.807 \pm 0.022$ ($\sigma$=0.172, 62 stars) in field A; and  
$<V>=19.320\pm 0.023$ ($\sigma$=0.159, 46 stars), $<B>=
19.680 \pm 0.024$ ($\sigma$=0.163, 46 stars) in field B. 
The $I$ light curve coverage is inferior, being based only on the 2001 data,
and we do not expand further on this band.

The double-mode RR Lyrae stars in our sample are slightly overluminous 
compared to the single-mode pulsators: 
$<V_{RRd}> =19.394\pm 0.066$ ($\sigma$=0.161, 6 stars), and 
$<V_{RRd}> =19.239\pm 0.085$ ($\sigma$=0.171, 4 stars), 
in field A and B respectively. A similar result
is found also by A96/A00.

There is a difference of 0.092 and 0.127 mag, respectively, between the
average  $V$ and $B$ magnitudes of the RR Lyrae stars  in the two fields,
explained by a difference  in the reddening towards the two areas, with field A
being   0.03 mag more reddened than field B (see Section 4). A few RR Lyrae's
were discarded when calculating the above average values. These were 11
objects  (4 RRc and 4 RRab in field A, and 2 RRc and 1 RRab in field B), for
which systematic  offsets were present between the  1999 and 2001 light curves,
possibly due to not resolved blends in either of the two datasets; variable
stars  whose light curves were   not evenly covered (2  stars in field A and 3
in field B, respectively); and  1 object in field A which fell on a bad CCD
column in the 1999 run.

The intrinsic dispersions of the average $B$ and $V$ apparent luminosities
(and, in turn, of the  average $B$ and $V$ absolute magnitudes) of the  RR
Lyrae's in the two fields are: $\sigma_B \sim 0.17$ and   $\sigma_V \sim 0.16$
mag, respectively. A number of sources  contribute to these dispersions:  
\begin{itemize} 

\item 
the internal photometric errors of  the present photometry: 0.02 mag standard
deviation of the  average

\item 
the metallicity distribution  of the RR Lyrae's in our sample.  The dispersion
around the mean value of [Fe/H]=$-1.48$ found from the VLT  low resolution
spectroscopy of 101 RR Lyrae  stars in the sample  is $\sigma$ = 0.29 dex (see
Section 3.3 and Gratton et al. 2000b). This metallicity dispersion well
compares with the $\sim$0.4 dex dispersion we would derive taking into account
the total period distribution spanned by the {\it ab} type RR Lyrae's in our
sample (0.$^d$40--0.$^d$74). According to $\Delta {\rm
M_V(RR)}/\Delta$[Fe/H]=0.2 mag/dex, these  metallicity dispersions correspond
to a magnitude dispersion of about 0.06 mag.

\item 
the level of evolution off the Zero Age Horizontal Branch (ZAHB) of the 
variables in  our sample. Here what matters
is not the global systematic difference between  ZAHB and average luminosity of
the RR Lyrae's,  but rather the dispersion  around the average  RR Lyrae
luminosity due to the evolution off the ZAHB of each individual RR  Lyrae.
Sandage (1990) studied the vertical height  of the HB of a
number of globular clusters of different metallicity,  and found for M3 and NGC
6981, clusters whose metallicity is close to the  average value of the LMC,
HB luminosities with standard dispersions of 0.064 and 0.097 mag
respectively, giving an average value of 0.08 mag\footnote{ This is likely to
be an overestimate of the real contribution due to evolution, because the field
RR Lyrae population should be dominated by stars  closer to the ZAHB than
variables in clusters at same metallicity  (see Carney et al. 1992)};  
and, finally 
\item the intrinsic depth of the  observed fields. 
\end{itemize}

Adding up in quadrature all dispersion contributions except the latter, we
obtain a total dispersion of about  0.10 mag to compare with the observed $V$ 
dispersion of 0.16 mag. This gives us some hint on the actual intrinsic depth
of our observed fields, for which we derive a most probable value of 0.13 and
an upper limit of about   0.15 mag corresponding to a dispersion in depth of
6.5\% and 7.5 \% or 3.3 kpc and 3.8 kpc (for an assumed distance  modulus
$\mu_{{\rm LMC}}$=18.50 mag, see Section 7).

Based on the spatial density profile of the MACHO RRab stars and on data for 
six additional fields from Kinman et al. (1991), A00 conclude that the 
majority of the old and metal-poor LMC field stars are likely to lie in a disk
with $i$=35 $\deg$, line  of nodes running North-South, and center near  the
optical center (Westerlund 1997), and not in a spheroid.  Our results on the
intrinsic depths of field A and B are not in  contrast with A00 results.

The comparison of the apparent luminosity of our RR Lyrae sample  with other
estimates available in the literature (Walker 1992a; Udalski et al. 2000,
Udalski 2000b, and A00) is postponed to Section 6. 

We have formed $<B>-<V>$ and $<V>-<I>$   colors from the intensity average $B,
V,$ and $I$ magnitudes for all the variables with complete light curves. These
intensity average magnitudes and colors ($<V>$, $<B> - <V>$,  $<I>$, $<V> -
<I>$) were used  to place the variables on the color-magnitude diagrams of the
two fields,   shown in Figure~4 and ~5.

\subsection{The period distribution}

Two peaks are clearly visible in the period distribution of the single-mode RR
Lyrae's in our sample (115 objects),  corresponding to the {\it c} (38
objects), and the {\it ab} (77 objects)  type pulsators (see Figure 8 of DF02).
The number of {\it c}-type  pulsators divides almost equally among the two
fields (20 and 18 RRc's in field A and B, respectively).  On the other hand,
the number of RRab's is about 50$\%$ larger in field A. This difference,
although not statistically significant, may suggest that we could be missing 
some {\it c} type pulsators in field A,  likely due to the smaller amplitude of
the variables  and the larger crowding of the field.

The mean period of the {\it c} and {\it ab} type RR Lyrae's is: 
$< {\rm P_{RR_c}} >$=0.$^d$327 ($\sigma$=0.047, 38 stars), and 
$< {\rm P_{RR_{\rm ab}}} >$=0.$^d$580 ($\sigma$=0.064, 77 stars), 
respectively, to be compared with 0.$^d$342 and 0.$^d$583 of A96.  

A96, on the basis of their $< {\rm P_{RR_{\rm ab}}} >$=0.$^d$583,  conclude
that  the preferred period of the {\it ab}-type variables of the LMC falls
between the periods of the  Galactic RR Lyrae stars of  Oosterhoff (1939) type
I (Oo I)  and II (Oo II),  but  it is actually closer to the Oo I cluster
periods  (being $<$P$_{{\rm RR_{\rm ab}}} >$=0.$^d$55, and 0.$^d$65 in the  Oo
I and II clusters, respectively). This finding is confirmed and strengthened by
our results.

The period-amplitude diagram of MACHO Bailey diagram sample (935 RRab's,  see
Section 4.2 of A00), is found to be very similar to the ridgeline defined by
the RRab's in M3, from which A00 infer  similarity of the mean metallicity of
the Bailey sample with M3. $B$ and $V$ amplitudes (A$_B$, A$_V$) were
calculated for all  the RR Lyrae's in our sample with full coverage of the
light curve as the difference  between maximum  and  minimum of the best
fitting models (see DF02), and have been used together with the newly derived 
periods to build period - amplitude diagrams. The overlap in the transition
region between {\it ab} and {\it c} type is  small  (6 objects) and   the
transition period between {\it c} and {\it ab} type occurs in our sample  at
P$_{\rm tr} \sim 0.^d$40,  while P$_{\rm tr}$ = 0.$^d$457,  in A96.   The
period - amplitude distributions of the LMC variables were  compared with the
relations defined by the  {\it ab} type RR Lyrae's in the globular clusters M3,
M15 and $\omega$ Cen. RR Lyrae's in field B seem to better follow the
amplitude-period relations of the variables in M3 and to belong to  the OoI
type. Variables in field A, instead, have pulsational properties  more
intermediate between the two Oostheroff types (see Figure 9 of DF02).

According to A97, nine double-mode pulsators were expected to fall in the
observed areas. We actually detected all of them and also found evidence  for
one possible additional RRd candidate not previously known from A97 : star
\#2249 (see Table 5 of DF02). 

\subsection{The metallicity distribution}

Metal abundances derived by Bragaglia et al. (2001) for a  first subset of 6 of
the double-mode pulsators in our sample  using Preston (1959) $\Delta$S
technique range from $-$1.09 to $-$1.78, with  an average value of
[Fe/H]=$-1.49 \pm 0.11$ (6 stars, rms=0.28 dex), on Clementini et al. (1995)
metallicity scale,  quite similar to Zinn \& West (1984) scale  (see Section 4
of Bragaglia et al. 2001).

Metallicities obtained from the spectroscopic survey  conducted in 2001 with
FORS at the VLT,  for 101 of the RR Lyrae stars discussed in  this paper range
from $\sim -0.5$ to $\sim -2.1$  with an average  metal abundance of [Fe/H] = 
$-1.48 \pm 0.03$ ($\sigma$=0.29, 101 stars, see Gratton  et al. 2002b). These
metallicities are tied to Harris (1996) metal abundances for the Galactic
globular clusters NGC 1851, NGC 3201, and M 68  that we used as calibrators. 
The values (available at http://www.physics.mcmaster.ca/Globular.html) we have
used are:  [Fe/H]=$-1.26$, $-1.48$, and $-2.06$, respectively, to compare to  
$-1.33$, $-1.56$, and $-2.09$ of Zinn \& West (1984). Thus they are on a
metallicity scale that,  on average, is 0.06 dex more metal rich than Zinn \&
West scale.

Photometric metallicities from the parameters of the Fourier  decomposition of
the $V$ light curves were also derived for 29 RRab's in our sample, 6 of which
do not have spectroscopic metal abundances, appling Jurcsik \& Kovacs (1996)
and Walker \& Kovacs (2001) techniques (see DF02, for details).  These metal
abundances are on the metallicity scale defined by Jurcsik (1995) which is, on
average, about 0.2 dex more metal rich than Harris (1996) scale. The average
metallicity of this subsample of stars is: [Fe/H]=$-1.27$ ($\sigma$=0.35, 29
stars), in good agreement with the average values derived from the above
spectroscopic analyses, once differences between  metallicity scales are taken
into account. The comparison between  photometric and spectroscopic abundances
can be found in Gratton et al. (2002b).

The spectroscopic abundances also well compare with both the range and the 
most frequent value of A96 spectroscopic analysis of 15 LMC RRab's,  thus
confirming both the  similarity with M3 suggested by A00, and the  existence of
RRab's very metal poor which,  according to A00, might trace a spheroid
population.

\section{The reddening of the LMC}

Accurate corrections for interstellar reddening are crucial for precise 
distance estimates. On the other hand,  there is strong evidence that the
reddening within the LMC is patchy and varies from one region to the other. 
This originates quite a lot of confusion in the various distance
determinations  to the LMC.

\subsection{Literature values}

Bessel (1991) provides a re-examination of the reddening of the LMC 
obtained by several independent methods and concludes that the foreground
$E(B-V)$ reddening probably lies between 0.04 and 0.09 mag, while the average 
reddening within the LMC is probably about 0.06 mag but with many 
regions with higher and lower than average reddening.

Walker (1992a) derives reddening values ranging from 0.03 to 0.18 mag 
for the 7 clusters he analyzes based on a combination of several independent 
methods and estimates: (i) Burnstein \& Heiles (1982) maps; (ii) the position 
of the 
Red Giant Branch in the color magnitude diagram of the clusters (the 
so called $(B-V)_{0,g}$ method); (iii) the pulsational properties of the 
variables is his sample [namely: Sturch (1966) method applied to the
{\it ab} type RR Lyrae at minimum light and with $E(B-V)$=0.00 mag for 
the reddening at the polar caps; and the colors of the edges of the 
instability strip defined by the RR Lyrae's in each individual cluster of 
his sample].

The reddening maps (based on COBE/DIRBE and IRAS data) of  Schlegel,
Finkbeiner, \& Davis (1998) give a foreground reddening, measured from dust 
emission in an annulus surrounding the LMC, of $E(B-V)$=0.075.  Schlegel et al.
find that their reddenings are  offset by $E(B-V)$= +0.02 mag with respect to
those provided by Burstein \& Heiles (1982), which are based on HI column
densities and deep galaxy counts\footnote{On average, the reddenings of 
Schlegel et al. agree quite well with those usually adopted for the  Galactic
globular clusters (see data set by Harris, 1996) for reddening  values smaller
than 0.10 mag; for more reddened clusters, the reddenings by Schlegel et al.
are systematically larger by 30\%.}.   The same maps give  upper limits of
$E(B-V)$=0.218 and 0.128 for field A and B, respectively. However Schlegel et
al. maps do not provide reliable  reddening for extended objects like the
Clouds or M31 close to their  centers.

Foreground and internal reddenings have also been estimated from $UBV$
photometry of individual early type stars by Oestreicher, Gochermann  \&
Schmidt-Kaler (1995) and Oestreicher \& Schmidt-Kaler (1996). The values for
the directions of our two fields are:  $E(B-V)$=0.056 and 0.062 (foreground),
0.150 and 0.140 (internal).  We may then derive lower and upper limits of the
reddening of $0.056 < E(B-V) < 0.206$ and $0.062 < E(B-V) < 0.202$  for field A
and B, respectively.  However, Zaritsky (1999) noted that reddenings derived
from early type stars  may be biased towards large values, likely because these
stars are preferentially located near star forming regions.

Udalski et al. (1999a)  published reddening values
determined along 84 lines of sights  for all of the 21 LMC fields observed with
OGLE-II. They used the red clump stars for mapping  the fluctuations of mean
reddening in their fields, treating the  mean I-band magnitude as the reference
brightness. Differences of the observed $I$-band magnitudes of the red clump
stars were  assumed as differences of the mean A$_I$ extiction and transformed
into  differences of $E(B-V)$ reddening assuming the  standard extinction curve
$E(B-V)$= A$_I$/1.96 by Schlegel et al. (1998). We note that this
procedure neglects possible systematic  differences  in the intrinsic mean
$I$-band luminosity of the LMC red clump  due to age or metallicity variations
from field to field  of the LMC. As discussed in Section 6.2, this may be a too
crude  assumption.  The zero-points of OGLE-II reddening maps were based  on
determinations around two LMC star clusters, namely NGC 1850 [$E(B-V)=0.15 \pm$
0.05 mag, based on $UBV$ photometry by Lee 1995], and NGC1835 [$E(B-V)=0.13
\pm$ 0.03 mag, based on colors of RR Lyr stars, Walker 1993]\footnote{We recall
however that  the instability strip of NGC 1835 dereddened according to this
$E(B-V)$ value appears too blue}; and on determinations based on OB-stars in
the  field of the eclipsing variable star HV2274 (Udalski et al.
1999a)\footnote{The value  of $E(B-V)=0.149\pm 0.015$ used by Udalski et al.
for this star has been questioned by Nelson et al. (2000), and Groenewegen \&
Salaris (2001), who suggested lower values of $E(B-V)=0.088\pm 0.025$\ and
$0.103\pm 0.007$,  respectively.}. These reddenings range from  $E(B-V)$=0.105
to 0.201 mag, with an average value of 0.143$\pm$0.017 mag. This average value
is respectively 0.042 and 0.017 mag lower than  the reddenings adopted for the
LMC RR Lyraes in  Udalski (1998a) and  Udalski et al (1999b).

Much smaller reddenings are obtained by considering individual Cepheids.
Reddenings for a total number of 80 individual Cepheids have been published by 
Caldwell \& Coulson (1986) and Gieren, Fouqu\'e \& Gomez (1998). These are the
values used by Laney \& Stobie (1994) and most other works on Cepheids. The 
two sets of reddenings are on the same system, and are coincident for the
majority of the stars in common.  These Cepheids have distances  from the
centers of our fields ranging from 0.66 up 7.8 degrees. In Table~1 we give
identification number and reddening value in both  Caldwell \& Coulson (1986)
and Gieren, Fouqu\'e \& Gomez (1998) systems, for a  subsample of Cepheids with
projected distances within two degrees from either of the centers of our
fields. An average of these reddening values gives $E(B-V)=0.067 \pm$0.006 mag
(7 stars) and $E(B-V)=0.067 \pm$0.023 mag (15 stars) for field A and B,
respectively. A weighted average of the  reddening estimates (with weights
proportional to the inverse square of the projected distance from our fields)
for all Cepheids whithin the two  samples yields mean reddening values of 
$E(B-V)$=0.070 for field A and 0.063 for field B. Note also that a star-by-star
comparison shows that the reddening for the Cepheids are on average $\sim
0.06$~mag smaller than those given by Schlegel et al.'s maps. 

Which of these different reddening scales should we adopt?

Independent indications for the reddening values in our fields   can be derived
directly from the  properties of  RR Lyrae variables in our sample, i.e.  i)
from the colors at minimum light of the {\it ab} type pulsators  through the so
called  Sturch's method (Sturch 1966); and   ii) from  the colors of the edges
of the instability strip defined by  the full sample of RR Lyrae stars.

Before applying these methods to our fields, we note that a direct comparison
suggests that field A is more reddened than field B. In fact:

\begin{itemize} 

\item on average, RR Lyrae's in field A are fainter by $\sim$0.09 mag in $V$,
and $\sim$0.13 mag in $B$, compared to the variables in field B  

\item average colors of the {\it ab}, and {\it c} and {\it d}-type RR Lyrae's 
in  field A are redder (by 0.036 and 0.016 mag, respectively) 
\end{itemize} 

We conclude that reddening is likely to be $0.03\pm 0.01$~mag larger in field 
A than in field B. Such a difference is not unconsistent with the literature results 
mentioned above, and will be used in the following discussion when combining
results from the two fields.

\subsection{Reddening from Sturch's method}

Sturch (1966) derives the reddening of {\it ab} type RR Lyrae's  from the
(magnitude-average) color at minimum light $(B-V)_{min}$ (phases 0.5-0.8), the
period P, and the metal abundance  [Fe/H] of the variables. The application of 
Sturch's method requires the knowledge of the metallicity   of each individual
RRab. 
We have used Sturch's method as described in Walker (1990, 1998), where
the reddening zero-point has been adjusted to give E(B$-$V)=0.0 mag at the
Galactic poles, and the [Fe/H] is that of Zinn \& West (1984) metallicity
scale, whereby 
$$ E(B-V) = (B-V)_{min} -0.24P -0.056[Fe/H]_{ZW}-0.336$$ 

Metal abundances are  available for 56 of the {\it ab}-type  RR Lyrae stars
from our VLT  spectroscopic study. They are  given in the upper portion of
Table~2 (column 5), along with  magnitude-average colors at minimum  light
(column 3) and periods (column 4) of the variables. These metallicities are on
Harris (1996) scale ( see Gratton et al. 2002b, and Section 3.3). Reddenings
for these stars have been calculated from the above relation after having
offset all metallicities by $-0.06$ dex, to put them on Zinn \& West (1984)
scale. For 6 additional objects (listed in the lower portion of Table~2), 
metal abundances on Jurcsik (1995) scale were derived from the Fourier 
decomposition of the $V$ light curves. We have used the relation provided by
Jurcsik (1995; [Fe/H]$_{Yurcsik}$= 1.431[Fe/H]$_{ZW}$+0.880) to transform 
metallicities to Zinn \& West scale. 

Derived $E(B-V)'s$ are given in column 6 of Table 2. The average reddening
value for the stars in field A is $E(B-V)=0.133 \pm 0.005$ ($\sigma$=0.031,
average on 37 stars),  discarding  star \#19711 which gives a negative
reddening value,  and star \#26525 wich deviates more tha 2.5 $\sigma$ from the
average. The average value for the stars in field B is $E(B-V)=0.115 \pm 0.007$
($\sigma$=0.036, average on 23 stars), again suggesting the existence of
differential  reddening between the two areas ($\sim$ 0.02 mag). 

Walker (1998) found that  Sturch's method gives  reddening values
systematically larger by 0.02-0.03 mag  when compared to other  reddening 
determination techniques. We have verified Walker's finding applying Sturch's 
method to some of the RR Lyrae stars in two of the clusters used as
spectroscopic calibrators, namely M68 and NGC1851. Brocato et al. (1994) have
published $B,V$ CCD photometry of RR Lyrae stars in M68. For three RRab's in
their sample (namely stars V9, V10 and V12)  Sturch's method gives
$E(B-V)_{V9}$=0.040, $E(B-V)_{V10}$=0.039, and  $E(B-V)_{V12}=-0.011$, and an
average value $E(B-V)$=0.04, discarding  V12 which gives a negative reddening.
This average reddening is to be compared  with literature values in the range
from 0.02 to 0.04 mag (see Brocato et al. 1994, and reference therein), and
with $E(B-V)$=0.035 from Schlegel et al. (1998) maps. For NGC1851 we have used
Walker (1998) photometry for the RRab's  V1, V6, V7, V11, V12, V16, V17 and
V22. The average reddening from  Sturch's method applied to these 8 stars is
$E(B-V)$=0.055 ($\sigma$=0.012, 8 stars), to  compare with $E(B-V)$=0.02 by
Harris (1996), and 0.061 mag from  Schlegel et al. maps.  The present results
confirm that Sturch's method may  overestimate reddening by 0.01-0.02 mag,
therefore we conclude  that the reddening values for field A and B derived with
Sturch's  method can be considered as  upper limits of the actual reddenings in
these areas. 

\subsection{Reddening from the colors of the instability strip}

Reddening in our fields  may be better  constrained by comparing the edges of
the instability strip defined by the RR Lyrae stars with those defined by
variables in globular clusters of known $E(B-V)$. This same procedure was used
by Walker (1992a) to estimate the reddening  value in some of the LMC globular
clusters.

Walker (1998) presents dereddened (magnitude-average) $B-V$ instability-strip 
boundary colors  from precise observations of 9 Galactic and LMC globular
clusters covering a range in metallicity  from --1.1 to  --2.2 dex. The
dereddened color of the blue edge is at $(B-V)_0=0.18  \pm 0.01$ mag, with no
discernable dependence on metallicity, while the color of the red edge shows a
shift of 0.04$\pm$0.02 mag dex$^{-1}$ with metal abundance. On the other hand,
since the RR Lyrae's in the LMC have period and amplitude distributions more
similar  to those of the M3 variables (see Section 3.2), we can directly
compare the  observed average colors of the LMC RR Lyrae's with those of the M3
ones. A further advantage of this choice is the very low reddening of M3,  with
$E(B-V)$ values in the range from 0.00 to 0.01 mag  (Ferraro et al. 1997,
Harris 1996,  Schlegel et al. 1998, Cacciari et al. 2003, in preparation).

Corwin \& Carney (2001)  have published accurate CCD $B, V$ light curves for 
more than 200 RR  Lyrae variables in M3. The magnitude-average $B-V$ color of
the  observed blue edge defined by Corwin et al. M3 RR Lyrae's is 0.184 mag.
This value compares extremely well with Walker (1998) dereddened  color of  the
instability strip blue edge $(B-V)_0$=0.18, thus confirming the very low
reddening suffered by  M3. The M3 observed red edge is  at  $B-V$ = 0.402 mag
(Corwin \& Carney, 2001).

In Table~3 we list (intensity and magnitude-average) $B-V$ colors  for the 5
bluest and the  5 reddest LMC variables in field A and B combined sample.
Colors of the variables in field A have been made bluer by 0.03 mag to  put
them on the same reddening system of field B.

The First Overtone Blue Edge  (FOBE) of the LMC RR Lyrae strip is very well
defined by the three bluest stars in the left hand portion of Table 3, namely
field B star \#2517 and field A  stars \# 2623  and \#2223, giving an average
color of the FOBE $<B-V>_{FOBE}$ = 0.245 ($\sigma$=0.011).

The Fundamental Red Edge (FRE) of the RR Lyrae instability strip   is more
difficult to define. Spectroscopic metal abundances  are available for 4 of the
5 stars at the FRE, namely stars: \#8094, \#28293 in field A; and \#7468 and
\#5589 in field B.  Colors were corrected to account for the  difference in
metallicity with M3  (assumed at [Fe/H]=$-1.66$ according to Zinn \& West 1984)
by applying the   shift   with metal abundance found by Walker (1998).  Final
adopted colors for the stars at the FRE are given in column 9 of  Table 3. We
used the 3 reddest stars in the right hand portion of Table~3,  namely field A
stars \#8094, \#28293, and \#7468 in field B to  obtain an average color of the
FRE of $<B-V>_{FRE}$ = 0.507 ($\sigma$=0.023).

Differences in color between the LMC and M3 FOBE and FRE are 0.061 and 0.105
mag, respectively. From the weighted average of  these differences, with the
FOBE having double weight, the  reddening in field B should be $0.076\pm 
0.016$\ larger than in M3 (i.e., larger than $E(B-V)=0.010\pm 0.007$ mag). The
reddening estimated  from the colors of the RR Lyrae instability strip edges is
then $E(B-V)=0.086\pm 0.017$\ for field B, and $E(B-V)=0.116\pm 0.017$\ for
field A. These values agree well with the estimates obtained by Sturch's method
if the 0.02-0.03 mag offset suggested by Walker (1999) is considered.

We verified that  RR Lyrae's in both field A and B,   once corrected for the
above reddening values,  are very well confined within the edges of the 
instability strip defined by Corwin \& Carney (2001) M3 variables, thus 
confirming the similarity of our LMC RR Lyrae sample to the OoI  cluster M3,
and giving support to the reddening values we determined  in the two fields. 

\section{Our results}

\subsection{Reddening and luminosity}

Hereinafter, we will adopt reddening values of 
${\rm E(B-V)}=0.116\pm 0.017$\ and $E(B-V)=0.086\pm 0.017$\ for field A and B,
respectively.  Reddening in field A is 0.028 mag smaller than derived by
Udalski et al.  (1999a) in the same area, and on average these new reddenings
are 0.04 mag smaller than OGLE-II average reddening for the LMC  (${\rm
E(B-V)}=0.143$,  Udalski 2000b), and 0.03 mag smaller than estimated from the
UBV photometry of early type stars (Oestreicher, Gochermann  \& Schmidt-Kaler
1995; and Oestreicher \& Schmidt-Kaler 1996). On the other hand, while
reddening in field B within the quoted  uncertainties agrees with the estimates
based on Cepheids (0.067$\pm$0.006), that in field A is clearly larger, and on 
average our values are  0.035 mag larger than reddenings  derived from
Cepheids. However, this result is not totally surprising since most of the
Cepheids with measured reddening lie in regions far from the LMC bar and none 
of them falls in our fields. We also note that since we are using a large
sample of $\sim 100$~objects  projected into the direction of the bar, we also
eliminated the possibility of systematic differences in the position of the
barycenter, that is well possible when considering a small number of objects
like e.g. the globular clusters. 

The absorption corrected intensity average magnitudes of the total sample 
of RR Lyrae's with full coverage of the light curves are 
$<V>_0=19.053\pm 0.021 \pm 0.053$ and 
$<B>_0=19.329\pm 0.023 \pm 0.070$;
average values of  each field were corrected for the corresponding reddening,
adopting  standard values for the selective absorption R$_V$=3.1 mag, 
R$_B$=4.1 mag (Cardelli et al. 1989).  The first error bar is the internal
dispersion of the average, while  the second term is due to the 0.017 mag
uncertainty in the reddening, still  by far the largest uncertainty source. The
final error should also include  the contributions of the uncertainty of the
photometric  calibration and of the aperture corrections (0.017 mag in $V$ and
0.032  mag in $B$, and   0.023 in V and 0.019 in B; see DF02).

Adding up in quadrature all error contributions  we derive for 108 objects:
$<V>_0=19.05\pm 0.06$ and $<B>_0=19.33\pm 0.08$  for the average dereddened
apparent luminosities  of our sample of RR Lyrae stars in the bar of the LMC.
A summary of our photometric results is provided in Table~4.

\subsection{The luminosity-metallicity relation}

The absolute magnitude of the RR Lyrae's, M$_{\rm V}$(RR), is known to depend
on  metallicity [Fe/H] according to the relation  M$_{\rm V}$(RR)=$\alpha
\times$ [Fe/H] + $\beta$,  but rather large uncertainties exist both on the
slope $\alpha$ and on the zero-point $\beta$ of this relation.  The zero-point
has been already discussed (see Section 1.1): we will now consider the slope.

The exact value of $\alpha$ has a large impact on several relevant 
astrophysical problems. For instance, when coupled with the almost constant 
value of $\Delta{\rm V}^{\rm HB}_{\rm TO}$ found for globular clusters,  the
M$_{\rm V}$(RR) - [Fe/H] relationship implies i) that all clusters are  more or
less coeval  if the high value of $\alpha$ is chosen, or, conversely, ii) that
there is a significant age-metallicity dependence in the family of the galactic
GCs (see Sandage 1993b and references therein) if a  shallow slope is the
correct value. The whole Galactic formation scenario is thus heavily affected
by the precise knowledge of $\alpha$.  

RR Lyrae stars in the LMC bar may  play a key r\^ole in the definition of
$\alpha$, since they can be considered all at the same distance from us, are
very numerous, and span more than 1.5 dex range in  metallicity. They can be
used to firmly constrain the value  of $\alpha$, independently of any model
assumption, the only limiting factor being the  intrinsic spread in the
luminosities of the RR Lyrae stars due to their evolution off the  ZAHB. To
minimize this effect a fairly large number of variables is  needed. 

We have combined the individual metal abundances of the RR Lyrae stars  derived
from our spectroscopic study with the corresponding dereddened mean  apparent
$V$ luminosities, and determined the slope of the luminosity-metallicity
relation for RR Lyrae stars. We used 85 out of the 101 RR Lyrae stars we
analyzed spectroscopically, i.e. those which have well sampled $V$ light curves
and no shifts between 1999 and 2001 photometries. They were divided into 5
metallicities bins; for each bin we  computed the average metal abundance, the
average dereddened apparent  luminosity $V_0$, and their root mean square
errors (see details in Gratton et al. 2002b).  A least square fit of these
average  values weighted by the errors in both variables gives the following
relation  for the apparent luminosity-metallicity relation of RR Lyrae stars:

$$<V_0(RR)> = [0.214 (\pm 0.047)] \times({\rm [Fe/H]} + 1.5) + 19.064(\pm 0.017)$$

where the error in the slope was evaluated via Monte Carlo simulations.

This relation is shown in Figure~6, where open and filled symbols are  used for
single and double-mode RR Lyrae stars, respectively. The mild slope we derive
is in very good agreement with recent results based on the HB luminosity of 19
globular clusters in M31: $\Delta  {\rm M_V(RR)}/\Delta$[Fe/H]= 0.22 mag/dex
(Rich et al. 2001) and it also agrees with the 0.20$\pm0.04$ slope found by the
Baade-Wesselink analysis of Milky Way field RR Lyrae stars: a confirmation that
the same  luminosity-metallicity relation for HB stars is valid  in three
different environments, namely M31, the LMC  and the Milky Way. Figure~6 does
not show any clear evidence for the  break in the slope around [Fe/H]=$-1.5$
suggested by recent  evolutionary/pulsation and HB models. 

Finally we note that in the $V_0~ vs ~{\rm [Fe/H]}$ relation the LMC RRd's  are
systematically offset to slightly higher luminosities, thus  giving support to
the hypothesis that the double-mode pulsators  may be more evolved than their
single-mode pulsator  counterparts.

That RRd's might be more evolved than single-mode RR Lyrae stars has often been
suggested on both observational and theoretical grounds,  however here for the
first time we can test this hypothesis  on a large sample of stars with known
metal abundance and all at same distance from us.

\section{Comparison with previous photometric results}

\subsection{The RR Lyrae}
 
The average absorption corrected apparent $V$ luminosity of the 108  RR Lyrae's
with full coverage of the light curve in our sample is   $<V(RR)>_0=19.06 \pm
0.02 \pm 0.03 \pm 0.05$ at an average metallicity of  [Fe/H]=$-1.5 \pm 0.03$,
on Harris (1996) metallicity scale  (see  Section 3.3 and Gratton et al.
2002b), and for  reddening values of 0.116$\pm 0.017$ and 0.086$\pm 0.017$ mag 
in field A and B, respectively (see Section 4). Here 0.02 mag is the standard
deviation of the average, 0.03 mag is the  contribution due to uncertainties of
photometric  calibration and aperture corrections (0.017 and 0.023 mag,
respectively),  and 0.05 mag is the  absorption contribution due to the 0.017
mag  uncertainty in the reddening.

This value can be compared with the following estimates available in the 
literature: (i) $<V(RR)>_0$=18.95$\pm $0.04, Walker (1992a; i.e. 19.04$\pm
0.04$ at [Fe/H]=$-1.5$), based on 182  variables in 7 LMC globular clusters at
[Fe/H]=$-$1.9 and for an average  reddening  value of $<E(B-V)>$=0.09 mag.  The
0.04 mag error bar is the internal error, and Walker   estimates that
systematic errors due to photometric zero-points and absorption are of the
order of 0.02--0.03 and 0.05 mag, respectively; (ii)  $<V(RR)>_0$=19.14,
derived from A00 median value of 680 RRab's  observed by the MACHO microlensing
experiment, at   [Fe/H]=$-$1.6  (i.e. 19.16 at [F/H]=$-1.5$) and for
$E(B-V)$=0.1 mag (Bessell 1991). A00 do not provide an evaluation of the
error,  and we have  worked them out as follows:  (a) we estimate a 0.073 mag
photometric zero-point uncertainty based on the  combination of A99: 0.021 mag
internal precision of the MACHO photometry, and A96: $\pm$ 0.07 mag mean error
of a single point of the RR Lyrae photometry; and (b) 0.06 absorption 
uncertainty based on a 0.02 uncertainty in the reddening. Finally, the 0.09 mag
total error is likely to be an underestimate since  it does not include the
standard deviation of the average (not provided by  A96 and A00). In this
respect we notice that A97 quotes  0.10 mag as a conservative estimate of the 
photometric uncertainties  of MACHO photometry for the RR Lyrae stars;  and
(iii) $<V(RR)>_0$=18.91 $\pm 0.01$ mag at [Fe/H]=$-$1.6  (i.e. 18.93 at
[Fe/H]=$-1.5$) and for an  assumed reddening value of 0.143 mag, in Udalski et
al. (2000b).

In Table~5 we summarize the $<V(RR)>_0$ values from the  literature transformed
to [Fe/H]$-$1.5, and we also list the associated  errors divided in internal
contribution and systematic errors (photometric  zero-point and absorption
components).

Summarizing: 

(a) The present average luminosity of the RR Lyrae stars in the  LMC bar agrees
well with Walker (1992a) photometry of RR Lyrae stars in  the LMC globular
clusters  (both including or discarding NCG 1841), thus ruling out the
existence  of  any intrinsic  differences between field and cluster variables.
Walker's and our $V$ photometries are both on the same consistent Johnson 
photometric system, and we used both reddenings values measured  with the very
same procedure  based on the intrinsic properties  of the stars whose average
luminosities are compared.

(b) Our $<V(RR)>_0$ is about 0.1 mag brighter than MACHO A00 value, contrary 
to A99  who found a  systematic negative shift of $\Delta V =-0.035$ between
MACHO and other photometric works,  implying higher luminosities for MACHO. As
discussed in Section 2.2 and DF02 this difference is almost interely due  to 
the difference existing between MACHO and our photometry for field A,  the
results for field B being in agreement with ours. On the other hand  the
average luminosity of the RRd stars in our sample:  $<V> =19.331\pm 0.055$
($\sigma$=0.175, 10 stars) is in  excellent agreement with A00 recalibration of
A97 luminosity of the LMC RRd's: $<V> =19.327\pm 0.021$.

(c) In Table 5 the two values listed for OGLE correspond to the average  
luminosity of the total sample of RR Lyrae stars (Udalski 2000b), and to the
luminosity of the subsample of RR Lyrae stars in  OGLE field LMC\_SC21  with
declination within the limits of our Field  A (Udalsky et al. 2000). Both OGLE
values are systematically  brighter by 0.14, 0.24 and 0.15 mag (or 0.12 mag if
NGC1841 is included)   than  the present photometry, A00 and Walker (1992a),
respectively. The total error bar of Udalski et al. (2000) determination  only 
marginally allows to make their $<V(RR)>_0$ values overlap with the  estimate
from the present photometry.  The offset between OGLE-II and our average
luminosity for the RR Lyrae stars  is due for about 0.06 mag to the zero-point
shift existing between the 2  photometries (caused by   overestimate of star
brightness at the faint limit of OGLE-II photometry,  likely due to the
different data reduction packages,  see DF02) and for the remaining 0.08 mag to
OGLE-II overestimate by about 0.02-0.03 mag of the reddening in these areas of
the LMC.

Using the average absolute magnitudes for RR Lyrae stars considered in Section
1.1 ($M_{V}=0.72 \pm 0.15$ at [Fe/H]=$-1.5$, average of the values  from
statistical parallaxes and the B-W method), we obtain a distance modulus of
$\mu_{LMC}$=18.34$\pm 0.15$. If we rather adopt the value recently  obtained
from the globular cluster Main Sequence Fitting by Gratton et al.  (2002a,
$M_{V}=0.61 \pm 0.07$  at [Fe/H]=$-1.5$), we  obtain a slightly longer distance
modulus of $\mu_{LMC}$=18.45$\pm 0.09$.

\subsection{ The clump }

We present here results on the clump, since it is relevant for the  distance to
the LMC. In fact,  the $I_0$ luminosity of the clump stars by the OGLE team
gives  the shortest distance modulus to the LMC. Udalski et al. (1998) modulus
of $18.08\pm 0.15$~mag has undergone a number of revisions which have moved it
towards longer values ($18.18\pm 0.06$: Udalski 1998b; $18.24\pm 0.08$: Udalski
2000a), but even the Popowski (2000)  latest increase to $18.27\pm 0.07$\ still
provides the shortest modulus derived so far for the LMC. Somewhat short
distance moduli were also provided by two LMC eclipsing  binary systems
(namely: HV2274, Guinan et al. 1998; and HV982: Fitzpatrick et al. 2000),
however   Ribas et al. (2002), in their recent analysis of a third LMC binary
system  (EROS 1044) conclude that the 3 binaries provide a consistent distance
modulus to the barycenter of the LMC of 18.4 $\pm$ 0.1.   

 A lively discussion is in place among scientists arguing  whether the clump
method is reliable or not, and invoking or not metallicity and age effects and 
dependences of the absolute luminosity of the clump at various different extent
(see e.g. Girardi \& Salaris 2001, and reference therein). A much longer clump
distance to the LMC has been published by Romaniello et al. (2000) based on HST
WFC2 observations of a field around SN1987A: $\mu_{LMC}$=18.59$\pm 0.04 \pm
0.08$.  More recent applications of this technique make use of  the $K$-band
luminosity of the LMC red clump stars (Sarajedini et al. 2002, Pietrzy\'nski \&
Gieren 2002).  The derived distance moduli are:  $\mu_{LMC}$=18.54$\pm 0.10$
and $\mu_{LMC}$=18.59$\pm 0.008 \pm 0.048$, respectively, and Pietrzy\'nski \&
Gieren (2002) emphasize that, due to uncertainties on population corrections,
actual errors of the clump  method are as large as  0.12 mag.

Hipparcos measured parallaxes for clump stars in the solar neighborhood and
derived for them an absolute V magnitude of about +0.8~mag, with a width of
about 0.7 mag (FWHM: Jimenez, Flynn \& Kotoneva 1998), and a $B-V$ color of
about 1.1 mag, with a  range of about  0.86$\lesssim B-V \lesssim$1.34 mag
(read from figure 4 of Jimenez et  al. paper).  This is 0.1 mag  {\bf fainter}
than the absolute V magnitude of RR Lyrae stars at [Fe/H]=$-1.5$,  the typical
value for the LMC variables, if the absolute magnitude given by the 
statistical parallaxes and the B-W method is adopted.

A00 found that the red HB clump of the 9M CMD is 0.28 mag {\bf brighter} than
the mean  brightness of the RR Lyrae stars. This difference is larger than
predicted  by theory for an old, coeval HB population, thus leading to the
conclusion that RR Lyrae and clump stars in the LMC have different ages.

We may directly compare the mean apparent $V$ magnitudes of the RR Lyrae and
clump  stars in our LMC fields. This comparison is shown in  Figure~7 where we
plot enlargements of the RR Lyrae and HB red clump region in the HR diagrams of
our fields. In each panel of the figure, the boxes outline  the clump stars of
the field, chosen to lie in the region $B-V$=0.65--1.20, and $V$=19.70--18.9 in
Field A, and $B-V$=0.62--1.17  and $V$=19.61--18.81 in field B, and containing
6728 and 3851 stars, respectively.  Gaussian fittings of the $B, V$, and $I$
luminosity functions   provide the following average magnitudes and colors of
the clump: 
$<B>=20.215$~mag, $\sigma=0.207$, 
$<V>=19.304$~mag, $\sigma=0.185$, 
$<I>=18.319$~mag, $\sigma=0.190$,
$<B>-<V>=0.911$~mag, and
$<V>-<I>=0.985$~mag
in field A; 
and 
$<B>=20.194$~mag, $\sigma=0.202$, 
$<V>=19.291$~mag, $\sigma=0.188$, 
$<I>=18.307$~mag, $\sigma=0.184$,
$<B>-<V>=0.903$~mag, and
$<V>-<I>=0.984$~mag in field B. 
These values can be compared with 
$<I_{clump}>$=18.25 reported by OGLE-II (from 
$<I_0>$=17.97  by Udalski 2000b, with  
$E(B-V)$=0.143, and A$_I$=1.96$E(B-V)$ from Udalski et al. 1999a).
A summary of our photometric results for the clump is provided in Table~5.

Clearly the LMC red clump  has a complex structure resulting from the
superposition of stellar populations with different masses and ages.
Evolutionary and age effects are then present and should be properly accounted
for when the HB red clump is used as a distance indicator. On the other hand,
the contribution to the clump by the old horizontal branch  in the LMC is very
well defined by its RR Lyrae stars. Figure~7 shows that  the average luminosity
of  RR Lyrae stars  corresponds to the lowest envelope of the clump. In fact,  
the average $V$ apparent luminosity of the clump stars is 0.108 and  0.029 mag
{\bf brighter} than the RR Lyrae level  in field A and B, respectively.  These
results show that: 
(1) the properties of the clump population in field A and B
are different, and a population gradient it is likely to exist between 
the two 
areas. 
(2) the properties of the clump population are 
quite different in the LMC and in the solar neighborhood. This is not 
unexpected. However the differences in ${\rm M_V}$\ are not those 
expected based on  OGLE-II results.

The fine structure of the red HB clump stars in our LMC fields will be
investigated more in detail in a following paper with the help of the
metallicity distribution drawn from about 300 clump stars we observed
spectroscopically. Here we only outline the effects on the clump distance
modulus to the LMC due to our new estimate for $I_0$.

According to our reddening estimates in the two fields, and adopting  Schlegel
et al. (1998)  value for the absorption in the $I$ band, A$_I$=1.94 $E(B-V)$,
the dereddened  average luminosity of the clump is $I_0$=18.12 $\pm$0.06 mag 
(where the error bar includes standard deviation of the average, photometric
zero-point,  and absorption contribution). This value is in excellent agreement
with Romaniello et al. (2000) but is 0.15 mag {\bf fainter} than in  Udalski
(2000b). This difference is accounted for by the 0.07 mag offset existing
between  our and OGLE-II photometries at the luminosity level of the clump 
(see Section 2.2 and DF02), and by a 0.08 mag difference in the I band
absorption, with OGLE-II reddening being about 0.04 mag larger than our
average  reddening (see Sections 4 and 5.1).

Using the metallicity-$I$ luminosity calibration for the clump stars by 
Udalski (2000a; M$_I$= 0.13([Fe/H]+0.25)$-$0.26), and assuming a mean 
metallicity of  [Fe/H]=$-$0.55  (again as in Udalski 2000a), we find a distance
modulus of  ${\rm (m-M)_{0,LMC}}=18.42\pm 0.07$; while if we adopt instead the
calibration  by Popowski  (2000; $M_I= 0.19$[Fe/H]$-$0.23) we obtain  ${\rm
(m-M)_{0,LMC}}=18.45\pm 0.07$. These values are larger  than those found by
Udalski (2000a,b) and Popowski (2000), and in  agreement  with almost all other
distance modulus determinations for the LMC  (see Figure~7).  Lacking at
present precise estimates of the metallicities of the clump stars in our
fields, these values have to be considered as preliminary. However,   given the
low sensitivity of M$_I$\ on metallicity, the assumption about  the metal
abundance of the clump stars has little impact on this distance  derivation.

\section{Summary and conclusions: the distance to the LMC}

We have presented results based on new $B, V, I$ photometry in the 
Johnson-Cousins  system  obtained for two fields  close to the bar of the LMC
and partially overlapping  with fields \#6 and \#13 of the MACHO microlensing
experiment.  125 RR Lyrae variables, 3 anomalous Cepheids, 11 Cepheids, 11
eclipsing binaries  and 1 $\delta$ Scuti star have been identified in the two
areas. The new photometry allows a very precise estimate of the average $V$ and
$B$  apparent luminosity of the RR Lyrae variables, as well as of the  $V, B,
I$ luminosity of the clump stars. The present new photometry has been
accurately compared with MACHO and OGLE-II photometries.  An estimate of the
reddening within the two fields was also obtained from  Sturch's (1966) method
and from the colors of the edges of the instability strip defined by the RR
Lyrae variables. This corresponds to $E(B-V)$=0.116, and 0.086 mag in field A
and B, respectively. These values are on average  about 0.03-0.04 mag smaller
than reddening derived from the UBV photometry of early type stars, and by
OGLE-II,  and about 0.035 mag larger than  reddening estimated from Cepheids
within two degrees from either of  the centers of our fields.

Spectroscopic metal abundances we derived for 80\% of the RR Lyrae stars  were
combined with individual dereddened average luminosities to determine the
luminosity-metallicity relation followed by the RR Lyrae stars:
$<V_0(RR)>=0.214({\rm [Fe/H]} + 1.5) +19.064$. The slope of this relation is in
very good agreement with recent results from GCs in M31 (Rich et al. 2001) and
shows no evidence for  a break around [Fe/H]=$-1.5$.

The average dereddened apparent luminosity of the RR Lyrae and clump stars  in
our LMC fields  is $<V(RR)>_0$=19.06$\pm$0.06 (at [Fe/H]=$-$1.5), and
$<V_{clump}>_0$=18.98$\pm$0.08,  where errors include standard deviation of the
average (0.02 mag and 0.06 mag for RR Lyrae and clump stars, respectively), 
uncertainty of the $V$ photometric calibration and aperture corrections, and
absorption contribution.

The present $<V(RR)>_0$   moves the B-W and  statistical  parallax
determinations of the distance modulus of the LMC  to
18.38$\pm$0.16\footnote{The B-W value would further move to 18.50 mag using the
new preliminary estimate of the magnitude of the Galactic field RR  Lyrae star
RR Cet:  M$_V$(RR)=0.56 mag at [Fe/H]=$-$1.5, from Cacciari et al. (2000) 
revision of the B-W method.} and 18.30$\pm$0.14, respectively. Using the most
recent results from the  Subdwarf Fitting Method (Gratton et al. 2000a), it is
18.45$\pm$0.09.

The present determination of the $I$ luminosity of the clump stars  corrected
for our new reddening  values leads to $<I>_0$=18.12$\pm$0.06 mag and moves the
clump  distance modulus of the LMC to 18.42$\pm$0.07 and 18.45$\pm$0.07 mag,
when  Udalski (2000a) or Popowski (2000) metallicity-$I$ luminosity relations
for the clump stars are adopted.

The state-of-the-art on the true distance modulus of the LMC as derived from 
various  techniques is summarized in  Figure~8, where different symbols are
adopted for Population I  (filled triangles) and II (filled circles)   distance
indicators.    Far for pretending to be exhaustive  (see Figure~8 and Table~10
of Benedict et al. 2002, for a more extensive  listing of extant LMC distance
modulus estimates),  Figure 8 is meant to  provide an overview of the results
from the most commonly used and more robust distance indicators found in the
LMC.  In panel (a) distance moduli from Pop. II indicators are based on Walker
(1992a) average luminosity of RR Lyrae's in the LMC Globular  clusters:
$<V(RR)>_0$=18.95 at [Fe/H]=$-$1.9 (transformed to  [Fe/H]=$-$1.5). Panel (a)
of Figure~8    well illustrates the dichotomy existing  between {\it short} and
{\it long} distance scales provided by the  different distance indicators: 
with a few exceptions (the red clump and the eclipsing binaries  method),  the
Population I distance indicators give a {\it long} distance modulus for the LMC
in the range from 18.5 to 18.7 mag, while  Population II  indicators based on
"field stars"  (Baade-Wesselink, and Statistical Parallax methods, in
particular) yield a  {\it short} distance modulus of 18.3, and indicators based
on "cluster stars",  as for instance the Main Sequence Fitting method ( MSF;
Gratton et al. 1997,  Carretta et al. 2000b)  and the globular clusters 
dynamical models, or the pulsational properties of RR Lyrae's in globular
clusters (Sandage 1993a) support a {\it  longer} modulus in the range from 18.4
to 18.6. The solid and dashed lines in panel (a) shows the HST key project on 
extragalactic distances preferred value $\mu_{{\rm LMC}}$=18.50 and its 
$1\sigma$=0.10 mag error bar (Freedman et al. 2001).

 The impact on the derived distance modulus of the LMC of  the present average
luminosities of RR Lyrae and clump stars in the LMC bar:  $<V(RR)>_0$=19.06 at
[Fe/H]=$-$1.5, and  $<I_{Clump}>_0$=18.12, and of our new reddening estimates:
$E(B-V)$=0.116 and 0.086 mag (in field A and B, respectively)  is shown  in
panel (b).  Further changes with respect to Figure~8a are i) the  revised
estimate of the distance to the LMC from the combinations  of the results for 3
LMC binary systems studied so far  (namely: HV 2274, HV982 and EROS 1044; Ribas
et al. 2002);  ii) the clump distance based on   K-luminosity (Sarajedini et
al. 2002), Pietrzy\'nski \& Gieren 2002);  iii) the revised HB luminosity from
the new Main Sequence Fitting (MSF) distances to NGC 6397, 6752 and 47 Tuc
based on ESO-VLT spectroscopy  of cluster giants and  turn-off stars (Gratton
et al. 2002a); iv) Cacciari et al. (2000) revised estimate of M$_V (RR)$  from
the Baade-Wesselink method; v) the use of A00 revised average  luminosity of
A97 LMC  RRd's: vi) the revision of the absolute luminosity  of RR Lyrae stars
from the white dwarf (WD) cooling sequence:  M$_V (RR)$=0.66$\pm 0.14$ at
[Fe/H]=$-1.5$ (Gratton et al. 2002a);  and vii) the use of the trigonometric
parallax of RR Lyr measured by HST (Benedict et al. 2002). Distance moduli
shown in Figure~8b are also listed in Table~6. Errors  for the Population II
indicators in Figure~8b and Table~6 include the  0.06 mag contribution of the
uncertainty in $<V(RR)>_0$.

A weighted average of the values in Table~6,  (with weights proportional to the
inverse square  of the error, and after having first weight-averaged   the
independent estimates available for the clump, the Miras, the B-W of RR Lyrae
stars, the RRd's, and the TRGB) shows that {\bf all} distance determinations
converge within 1 $\sigma$  on a distance modulus of $\mu_{\rm LMC}$=18.515$
\pm 0.085$ mag (where the  error bar is the 1 $\sigma$ scatter of the average)
thus allowing to {\bf fully  reconcile the long and the short distance scale to
the LMC}. This value is shown by the heavy long dashed line of Figure~8b  while
heavy dashed lines give the 1$\sigma$=0.085 mag errorbars. A straight average
of the values in Table~6, with no weights, would give $\mu_{\rm LMC}$=18.50$
\pm 0.10$.  We caution that distance moduli from Cepheids may be overestimated
by 0.06-0.09 mag due to an underestimate of the reddening adopted for Cepheids
by 0.02-0.03 mag.

\acknowledgments{
We thank the referee A. Walker for making constructive comments  and
suggestions concerning  the original manuscript which have definitely helped to
improve and  strenghten the paper, and  M. Feast for his comments on the first
version of the paper circulated as astro-ph/0007471. Special thanks go to P.
Montegriffo for  the development of the specific software used for the 
detection of the variable stars and in the study of their periodicities.  G.C.
whishes to thank M. Marconi and C. Cacciari for very helpful discussions about
the classification, according to Bailey types, of some of the RR Lyrae
variables, and about the use  of the Fourier decomposition parameters, and the
reddening of M3.

This paper utilizes public domain data obtained by the MACHO Project, jointly
funded by the US Department of Energy through the University of California,
Lawrence Livermore National Laboratory under contract No. W-7405-Eng-48, by the
National Science Foundation through the Center for Particle Astrophysics of the
University of California under cooperative agreement AST-8809616, and by the
Mount Stromlo and Siding Spring Observatory, part of the Australian National
University.

This work was partially supported by MURST - Cofin98 under the project "Stellar
Evolution" , and by MURST - Cofin00 under the project "Stellar observables  of
cosmological relevance".

\newpage
\figcaption[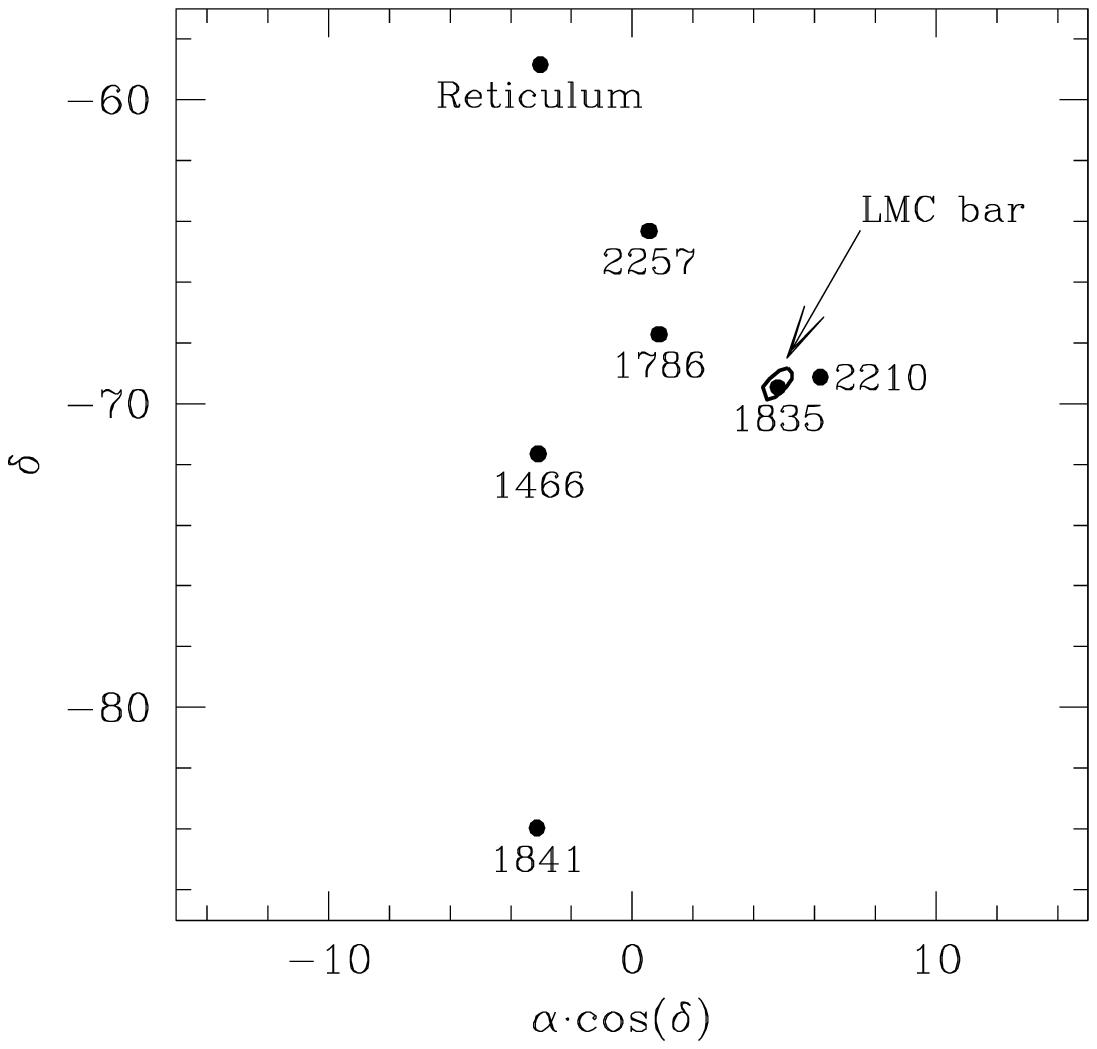]{Positions of the 7 LMC globular clusters 
studied by Walker (1992a) with respect to the LMC bar.}

\figcaption[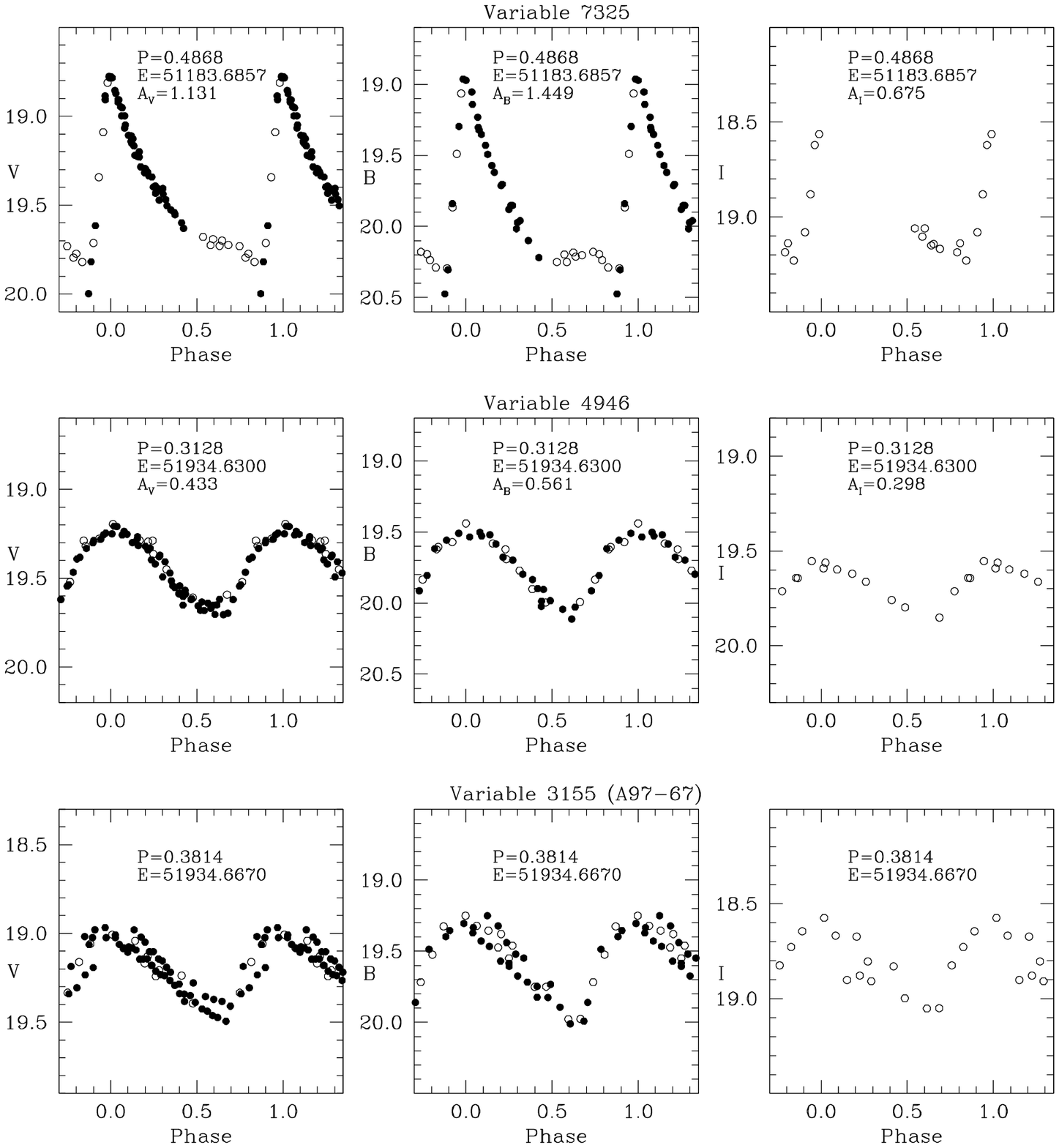]{$V, B, I$ light curves of RR Lyrae stars
falling 
in our fields. From top to bottom an RRab, an RRc and 
an RRd variable. Filled and open symbols are used for the 1999 and 2001 
data, respectively.} 

\figcaption[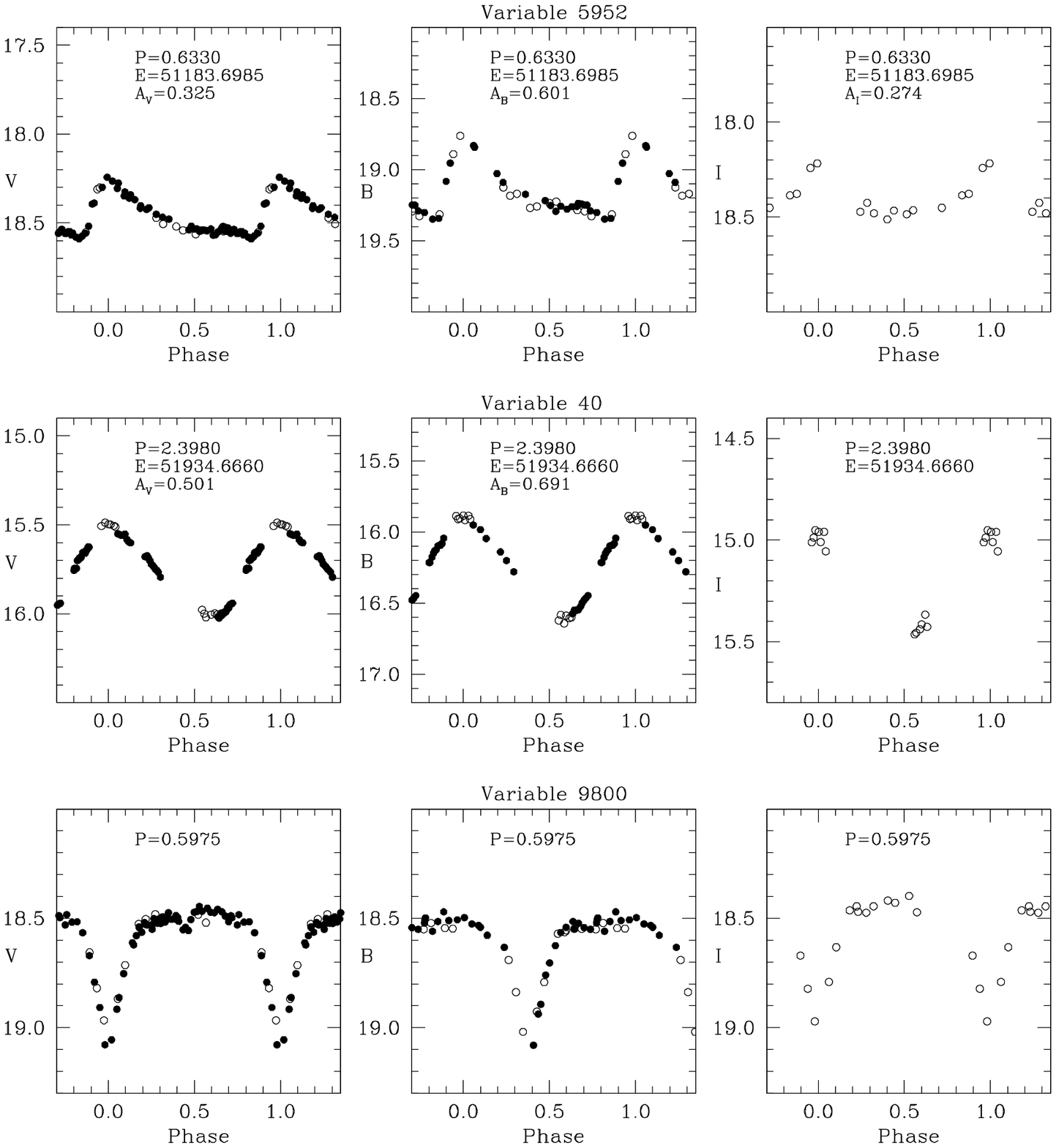]{
$V, B, I$ light curves of an anomalous Cepheid, a
classical Cepheid and an eclipsing binary falling in our fields. 
Different symbols are as in Figure~2.} 

\figcaption[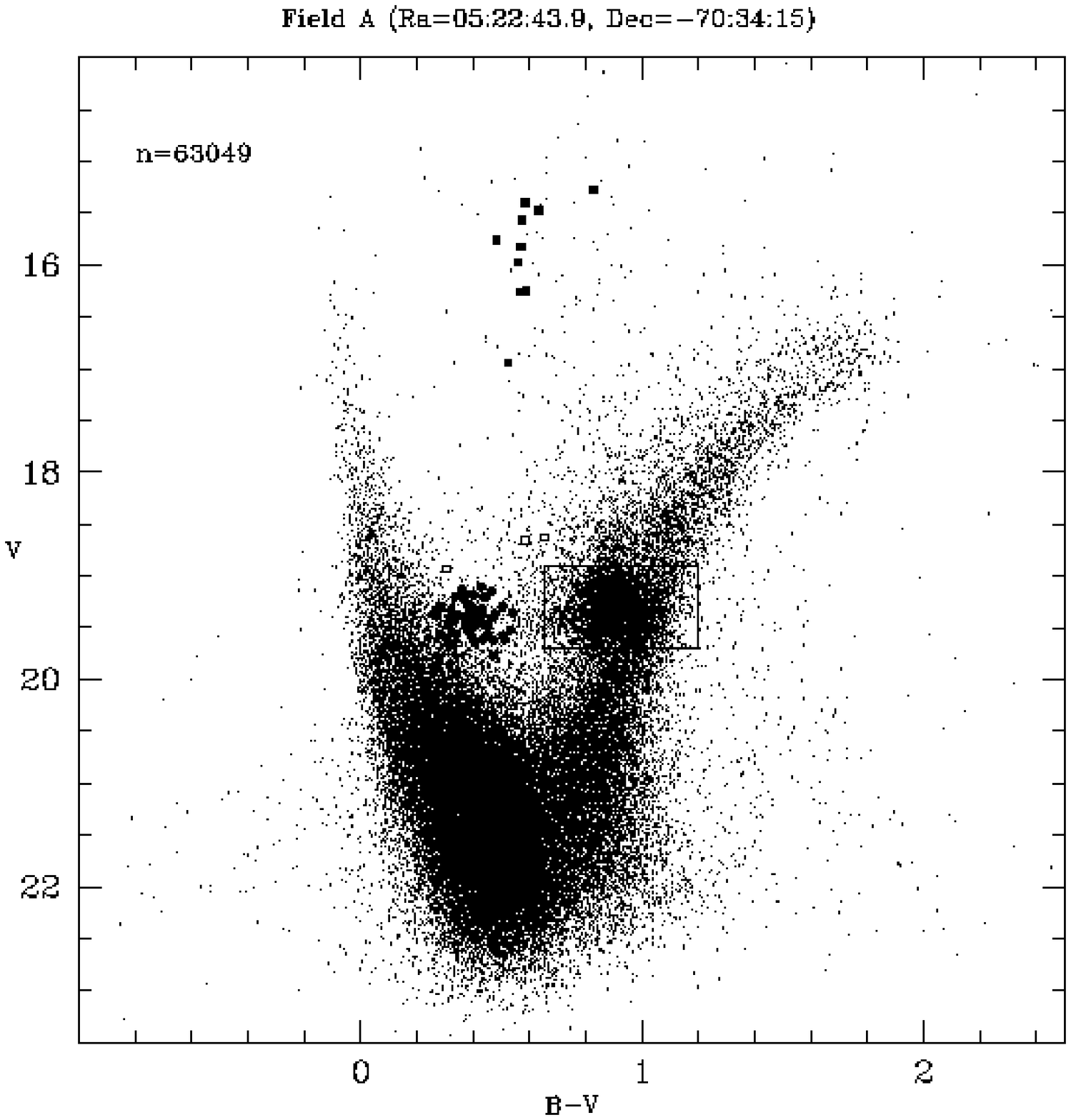,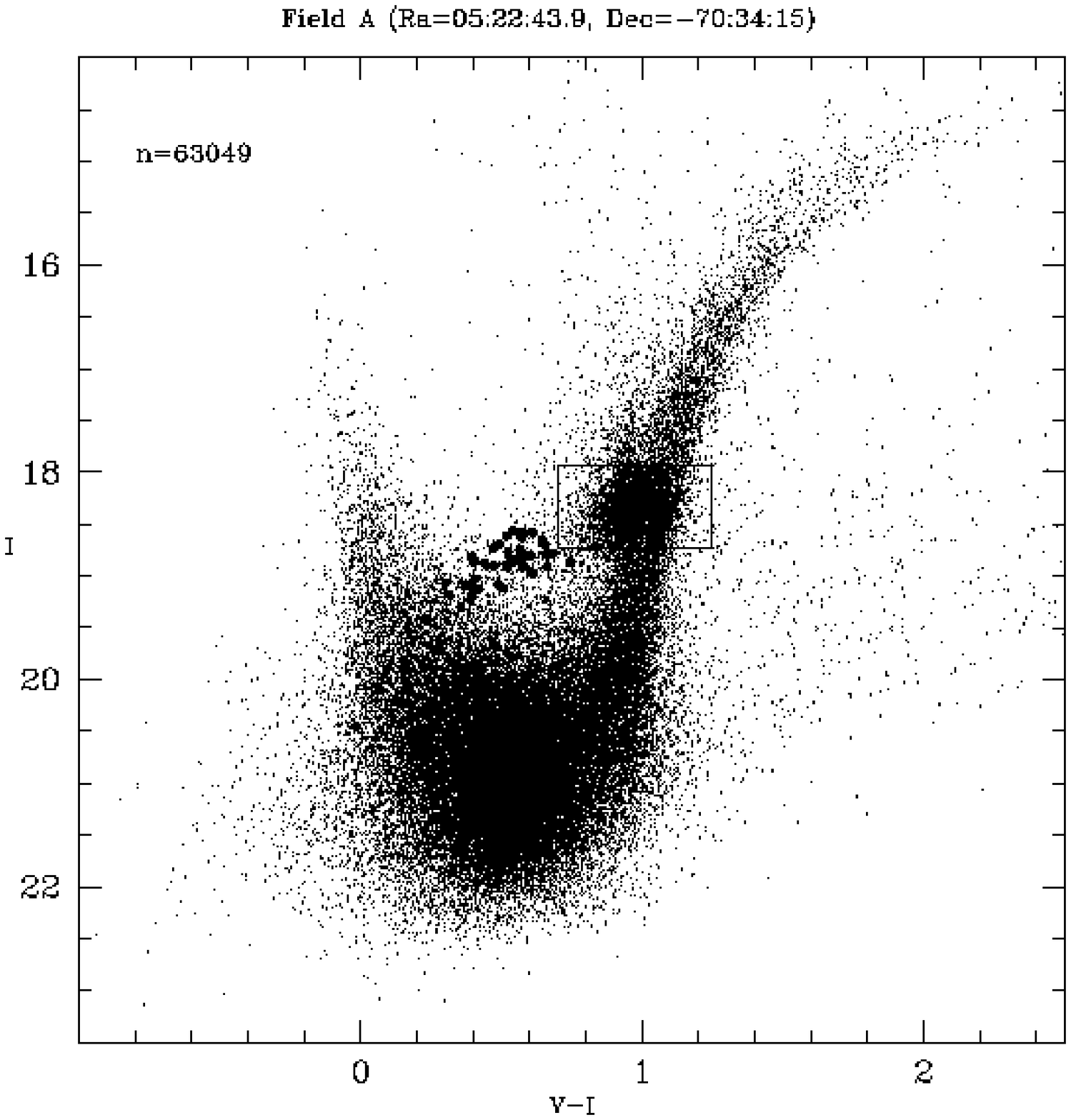]{ $V$ {\it vs} $B-V$, and
$I$ {\it vs} 
$V-I$ color - magnitude diagrams of field A from the ALLFRAME reduction
of the 2001 data. 
The box outlines the clump stars of this field.
Different symbols mark the variables identified in this 
field
(RR Lyrae stars: filled circles; anomalous Cepheids: open squares; Cepheids: 
filled squares; binaries: filled triangles; crosses: $\delta$ Scuti) 
which are 
plotted according to their intensity average magnitudes and colors.
Only the RR Lyrae stars are shown in the $I$ {\it vs} $V-I$ diagram.}

\figcaption[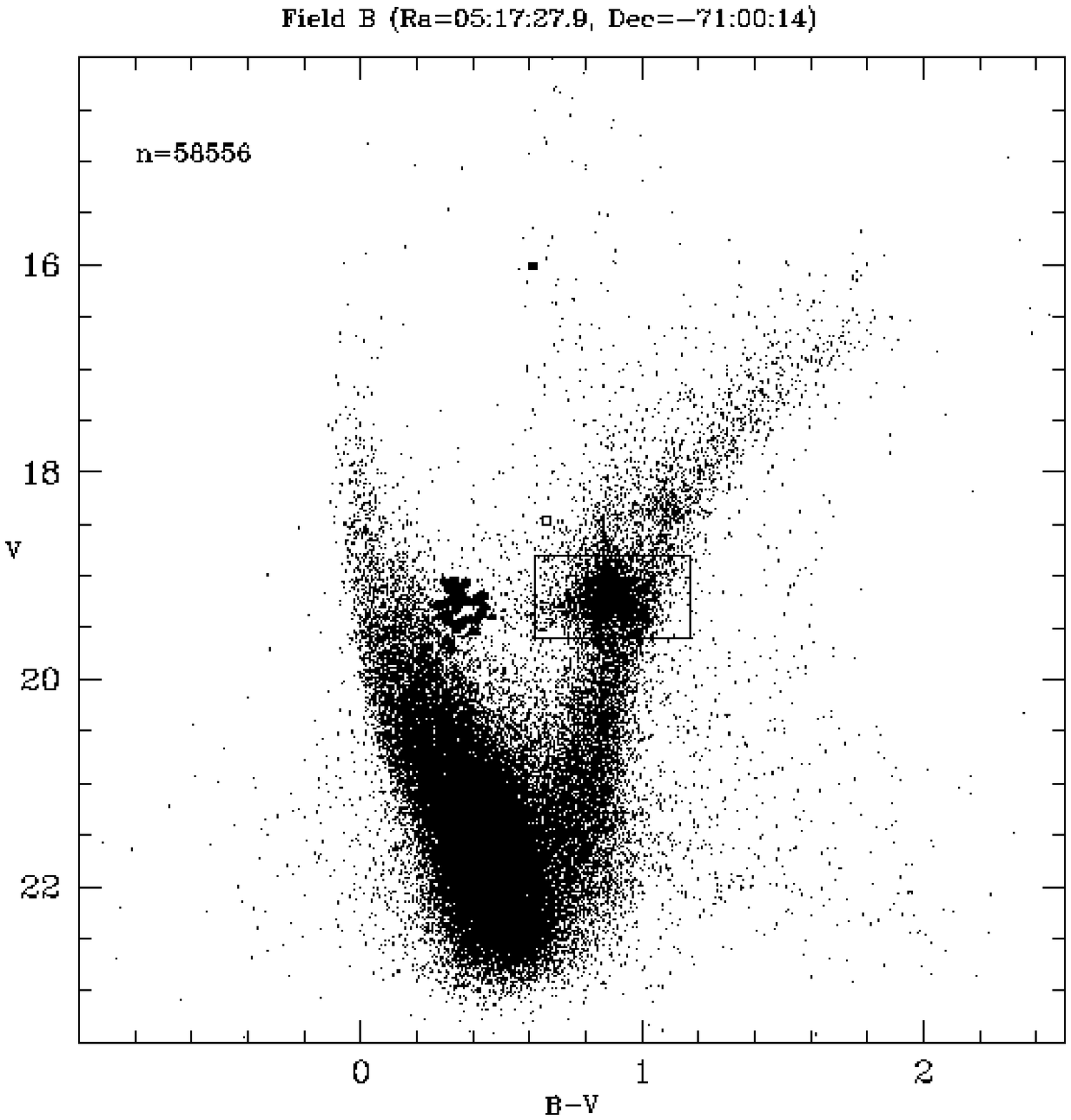,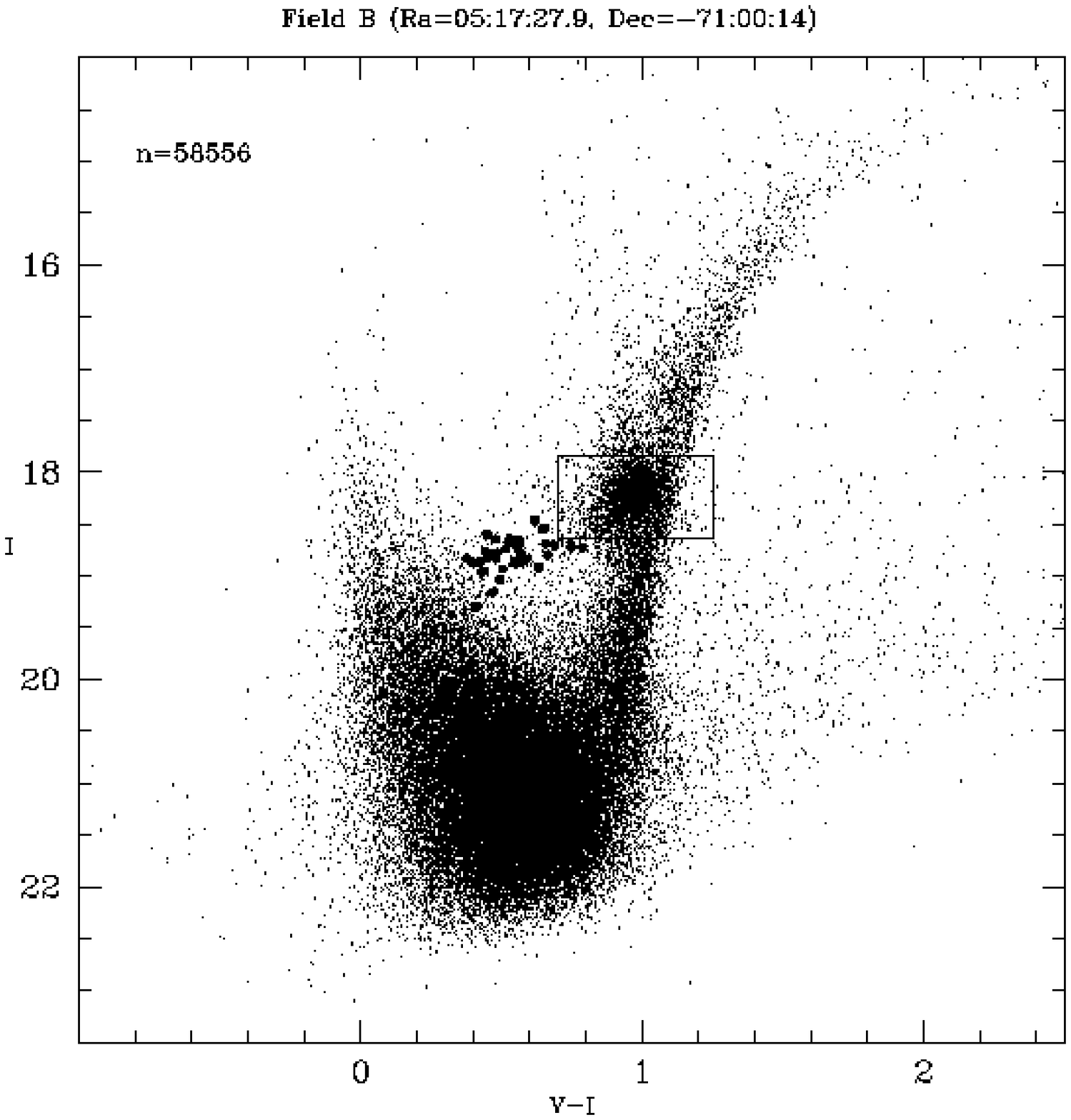]{Same as Figure~4 for field B.}

\figcaption[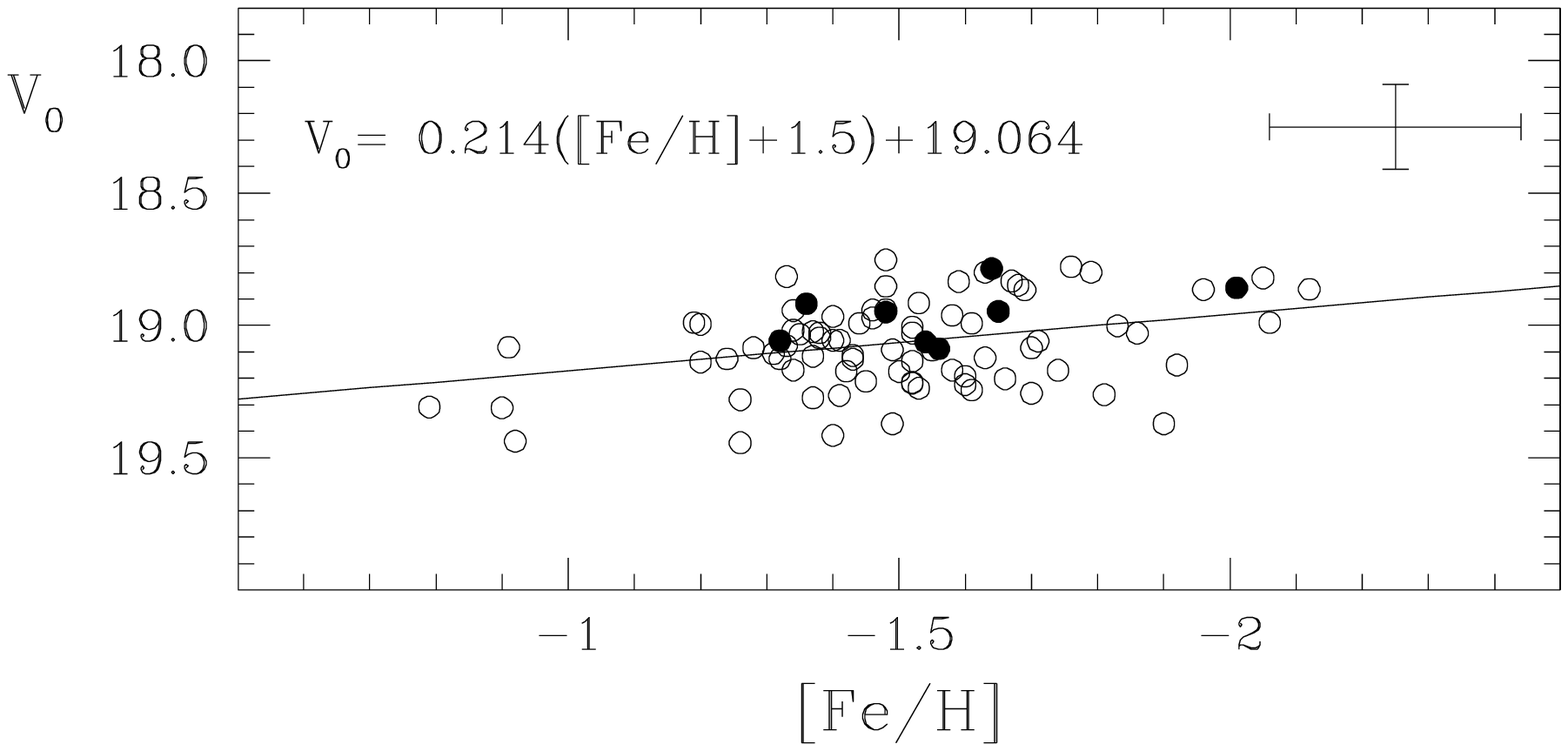]{The luminosity-metallicity relation defined by our
RR Lyrae stars. Filled circles mark the double-mode pulsators.} 

\figcaption[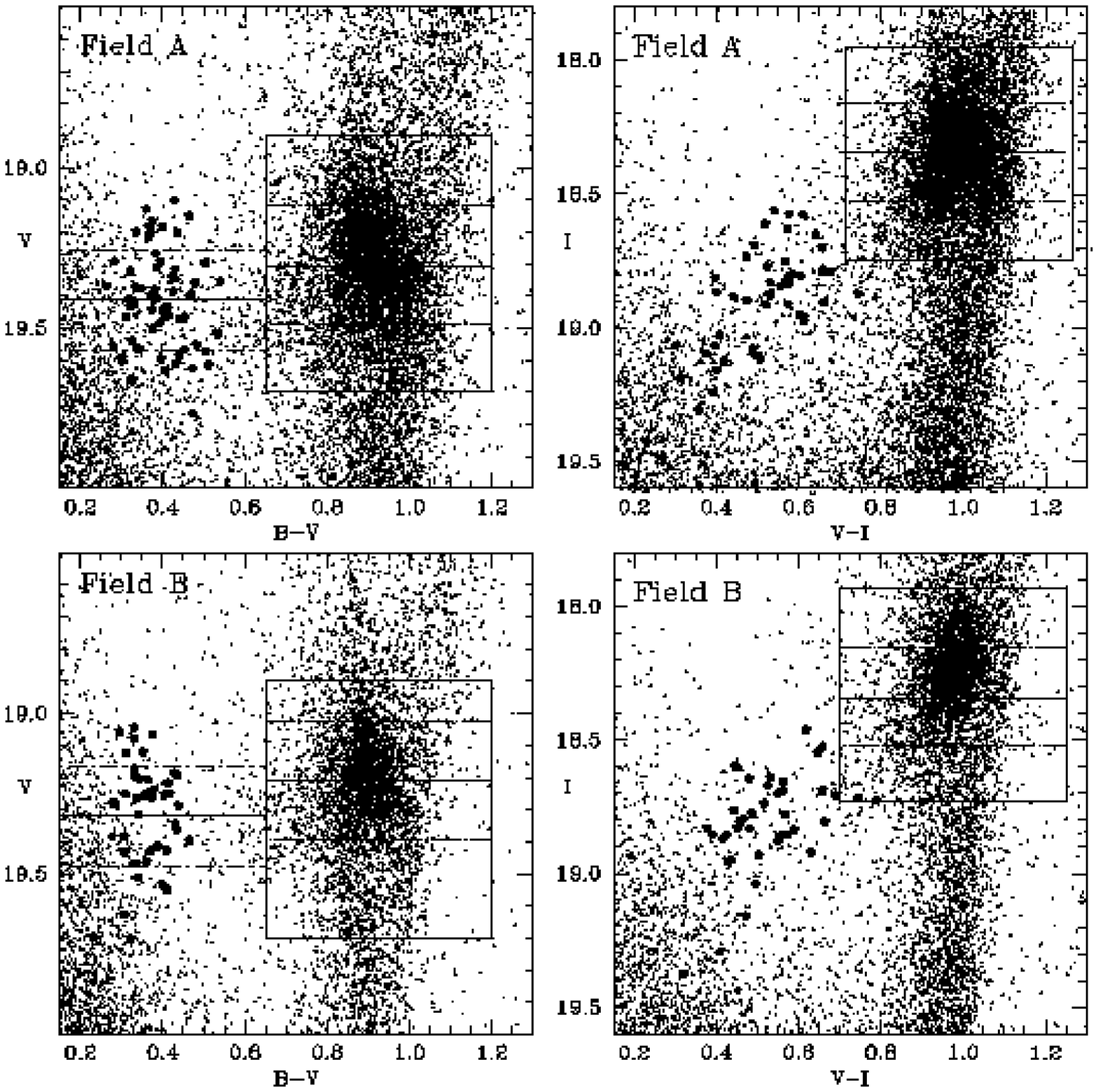]{Enlargement of the regions containing the
RR Lyrae and  clump stars in the  $V$ {\it vs} $B-V$, and  $I$ {\it vs} $V-I$
CMDs of field  A (top panels) and B (lower panels),  respectively. Filled
pentagons are the RR Lyraes plotted according to their intersity average
magnitudes and colors. Solid lines in the left hand panels indicate the average
levels of the  RR Lyrae's  ($<V(RR)_{A}>$=19.412 mag, $<V(RR)_{B}>$=19.320
mag)  and clump stars ($<V_{Clump A}>$=19.304 mag,  $<V_{Clump B}>$=19.291
mag)  in each field. Dotted-dashed lines represent  the 1 $\sigma$ deviations
from the averages ($\sim$0.16 and $\sim$0.19 mag for RR  Lyrae's and clumps
stars,  respectively).} 

\figcaption[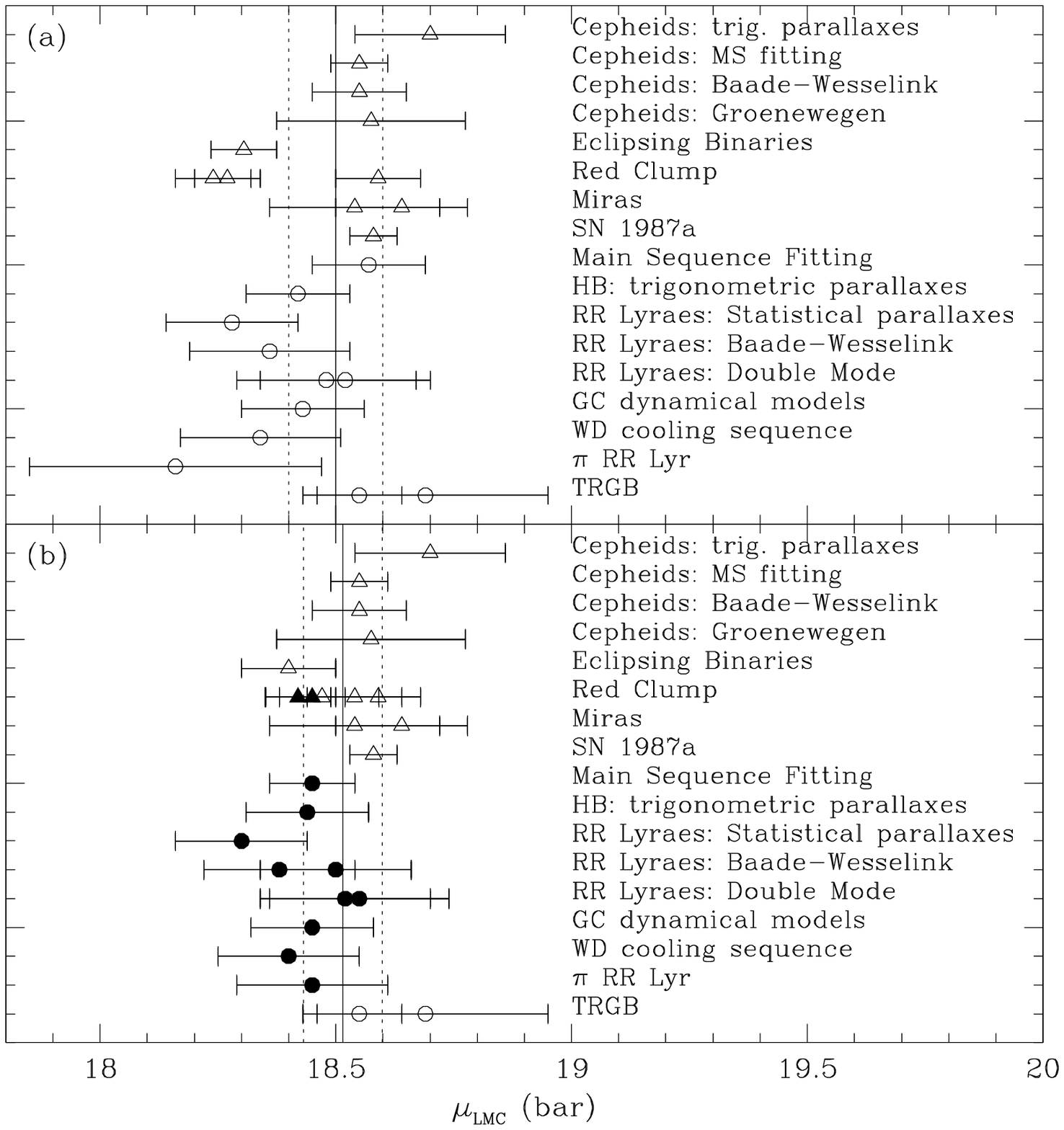]{The distance modulus of the Large Magellanic
Cloud  as derived from various distance indicators. Triangles and circles
mark   the Population I and II indicators,  respectively. In panel (a) distance
moduli from Pop. II indicators are based on Walker (1992a) average luminosity
of RR Lyrae's in the LMC Globular  clusters: $<V(RR)>_0$=18.95 at [Fe/H]=$-$1.9
(transformed to  [Fe/H]=$-$1.5);  while in panel (b)  distance moduli are based
on our new  average  luminosities of RR Lyrae and clump stars in the bar of the
LMC: $<V(RR)>_0$=19.06 at [Fe/H]=$-$1.5, and  $<I_{Clump}>_0$=18.12, and on our
new reddening estimates: $E(B-V)$=0.116 and 0.086 mag (in field A and B,
respectively). The solid and dashed lines in panel (a) shows the HST key
project on  extragalactic distances preferred value $\mu_{{\rm LMC}}$=18.50 and
its  $1\sigma$=0.10 mag errorbar (see Freedman et al. 2001, and  Section 7).
Data in panel (a) of the figure for the Cepheids are taken  from: (a)
trigonometric parallaxes, Feast \& Catchpole (1997; with errorbar revised
according to  Carretta et al. 2000b); (b) MS Fitting, Laney \& Stobie (1994 
value corrected to account for Hipparcos new distance modulus for the  Hyades,
and with errorbar revised according  to Carretta et al. 2000b); (c) B-W,
average of Gieren et al. (1998) value  corrected according to Carretta et al.
(2000b), and of Di  Benedetto (1997) estimate;  (d) P/L relation from
Groenewegen (2000); for the eclipsing binaries:  average from Guinan et al.
(1998) and Fitzpatrick et al. (2000); for the clump from: Udalski (2000a), 
Popowski (2000), and Romaniello et al. (2000); for  the Miras from: Van Leeuwen
et al. (1997) and  Feast (2000); for SN1987a from: Panagia  (1998); for the MSF
and the HB  stars from: Carretta et al. (2000b); for the RR Lyrae Statistical
parallaxes and B-W: from  M$_{\rm V}$(RR)=0.76 and 0.68 mag, respectively, and
Walker (1992a)  $<V(RR)>_0$; for the LMC RRd's from A97 and from Kovacs (2000);
for  the GCs dynamical  models from: Carretta et al. (2000b) revision of
Chaboyer et al. (1998);  for the white dwarfs cooling sequence  from: Carretta
et al. (2000b);  for the Hipparcos trigonometric parallax of RR Lyrae itself:
Perryman et al. (1997); and for the RGB Tip Cioni et al. (2000), and Romaniello
et al. (2000).  The revision of the LMC distance moduli due to  the present new
photometry and reddening is shown  in panel (b). Further changes with respect
to panel (a) are: (i) Ribas   et al. (2002) revised estimate of the distance to
the LMC from the 3 eclipsing  binary systems HV 2274, HV 982 and EROS 1044;
(ii) the clump distances based on the   K-luminosity (Sarajedeni et al. 2002,
and  Pietrzy\'nski \& Gieren 2002);  (iii) the revised MSF distances to GCs  by
Gratton et al. (2002a); (iv) Cacciari et al. (2000) revised estimate of M$_V
(RR)$  from the B-W method; (v) A00 revised average luminosity of A97 LMC
RRd's; (vi) the revised distance from the WD cooling sequence by Gratton et al.
(2002a);  and (vii) Benedict et al. (2002)  trigonometric parallax of RR Lyr
measured by the HST. The heavy long dashed line shows the average value of all
distance  moduli in panel (b):  $\mu_{\rm LMC}$=18.515 mag,  while heavy dashed
lines give the 1$\sigma$=0.085 mag errorbars (see Section 7).}

\begin{table}[ht]
\vskip 4 cm
\caption{Reddening values for Cepheids within two degrees from the centers
of either of our fields}
\vspace{0.5cm}
\begin{tabular}{ccccc}
\tableline
\tableline
\multicolumn{1}{c}{HV number} &\multicolumn{1}{c}{$E(B-V)$}&\multicolumn{1}{c}
{$E(B-V)$}&\multicolumn{1}{c}{d(A)}&\multicolumn{1}{c}{d(B)}\\
\multicolumn{1}{c}{}&\multicolumn{1}{c}{(1)}&\multicolumn{1}{c}{(2)}&
\multicolumn{1}{c}{(deg)}&\multicolumn{1}{c}{(deg)}\\
\tableline
12823& 0.058&      & 0.93& 1.30\\
~2549& 0.058&      & 1.13& 0.99\\
~2527& 0.070&      & 1.18& 0.66\\
12797& 0.070&      & 1.57& 1.09\\
~2694& 0.070&      & 1.67& 2.05\\
~6065& 0.070&      & 1.82& 3.11\\
~2749& 0.070&      & 1.92& 2.49\\
~2405& 0.070&      & 2.15& 1.54\\
~~909& 0.058&      & 2.66& 0.88\\
~2352& 0.100& 0.105& 2.71& 0.96\\
~2338& 0.040& 0.036& 2.74& 0.96\\
~~902& 0.070&      & 2.82& 1.02\\
~~900& 0.058&      & 3.03& 1.33\\
~~914& 0.070&      & 3.17& 1.95\\
~~881& 0.030&      & 3.56& 1.73\\
~2257& 0.060& 0.061& 3.74& 1.95\\
~~873& 0.130&      & 3.75& 1.92\\
~~878& 0.058&      & 3.76& 1.97\\
\tableline
\end{tabular}
\label{t:tab1}
\medskip

(1) Gieren et al. (1998); (2) Caldwell \& Coulson (1986)
\label{t:tab1}
\end{table}

\noindent
\begin{table}[ht]
\begin{center}
\caption{Individual reddenings 
from Sturch's method applied to the RRab stars with known metal
abundance.
Metal abundances in the upper portion of the table
derive from our spectroscopic study and are on Harris (1996) 
metallicity scale (see Section 3.3 and Gratton et al. 2002b).
Metal abundances in the lower portion of the table were derived 
from the Fourier parameters of the light curve and are on 
Jurcsik (1995) metallicity scale.
Reddenings were calculated using the relation 
given in Section 4.2, after having transformed metallicities to 
Zinn \& West (1984) scale.}
\vspace{0.5cm}
\scriptsize
\begin{tabular}{rccclc}
\tableline
\tableline
Id.~~&Field&   $<B-V>_{min}$         & $P$ &[Fe/H]&$E(B-V)$\\ 
   &     &mag.& day &     &  \\
\tableline
 2525& A & 0.535 &0.6161&$-2.06$ &0.169\\
 2767& A & 0.528 &0.5311&$-1.37$ &0.144\\
 3061& A & 0.496 &0.4762&$-1.26$ &0.119\\
 3948& A & 0.564 &0.6666&$-1.46$ &0.153\\
 4933& A & 0.559 &0.6135&$-1.48$ &0.162\\
 4974& A & 0.513 &0.5807&$-1.35$ &0.116\\
 5589& A & 0.555 &0.6365&$-1.60$ &0.159\\
 6398& A & 0.532 &0.5603&$-1.40$ &0.143\\
 6426& A & 0.540 &0.6622&$-1.59$ &0.137\\
 7247& A & 0.478 &0.5617&$-1.41$ &0.088\\
 7325& A & 0.490 &0.4868&$-1.28$ &0.112\\
 7468& A & 0.584 &0.6355&$-1.41$ &0.177\\
 7477& A & 0.527 &0.6564&$-1.67$ &0.130\\
 7609& A & 0.511 &0.5734&$-1.58$ &0.129\\
 7734& A & 0.538 &0.6170&$-1.39$ &0.135\\
 8094& A & 0.595 &0.7457&$-1.83$ &0.186\\
 8720& A & 0.479 &0.6577&$-1.76$ &0.087\\
 8788& A & 0.516 &0.5596&$-1.55$ &0.135\\
 9154& A & 0.508 &0.6198&$-1.66$ &0.119\\
 9245& A & 0.448 &0.5598&$-1.42$ &0.060\\
 9494& A & 0.518 &0.5786&$-1.69$ &0.141\\
10214& A & 0.500 &0.5999&$-1.48$ &0.106\\
10487& A & 0.531 &0.5896&$-1.52$ &0.142\\
12896& A & 0.574 &0.5737&$-1.53$ &0.189\\
15387& A & 0.512 &0.5598&$-1.81$ &0.146\\
16249& A & 0.480 &0.6048&$-1.86$ &0.106\\
18314& A & 0.526 &0.5871&$-1.71$ &0.148\\
19450& A & 0.537 &0.3979&$-0.90$ &0.159\\
19711& A & 0.361 &0.5530&$-1.68$ &$-0.010$\\
25301& A & 0.576 &0.5606&$-1.40$ &0.187\\
\tableline
\end{tabular}
\end{center}
\normalsize
\label{t:tab2}
\end{table}

\noindent
\begin{table}[ht]
\vskip 4 cm
\begin{center}
\vspace{0.5cm}
\scriptsize
\begin{tabular}{rccccc}
\multicolumn{6}{c}{Table~2 continued}\\
\tableline
\tableline
Id.~~&Field&   $<B-V>_{min}$         & $P$ &[Fe/H]&$E(B-V)$\\ 
   &     &mag.& day &     &  \\
\tableline
25362& A & 0.509 &0.5775&$-1.49$ &0.121\\
25510& A & 0.525 &0.6496&$-1.63$ &0.127\\
26525& A & 0.601 &0.5229&$-1.63$ &0.234\\
26821& A & 0.554 &0.5875&$-1.37$ &0.157\\
26933& A & 0.456 &0.4883&$-1.34$ &0.081\\
28293& A & 0.483 &0.6615&$-1.74$ &0.089\\
 1408& B & 0.453 &0.6260&$-1.70$ &0.065\\
 1575& B & 0.541 &0.6739&$-1.61$ &0.136\\
 1907& B & 0.524 &0.5805&$-1.52$ &0.137\\
 2249& B & 0.526 &0.6104&$-1.56$ &0.134\\
 2884& B & 0.515 &0.6194&$-1.90$ &0.140\\
 3400& B & 0.504 &0.4862&$-1.45$ &0.136\\
 4780& B & 0.576 &0.6176&$-1.20$ &0.162\\
 4859& B & 0.486 &0.5234&$-1.44$ &0.108\\
 5902& B & 0.459 &0.5698&$-2.12$ &0.108\\
 6440& B & 0.407 &0.4948&$-1.19$ &0.032\\
 6798& B & 0.498 &0.5841&$-1.20$ &0.092\\
 7063& B & 0.516 &0.6543&$-1.49$ &0.109\\
 7442& B & 0.519 &0.5780&$-1.58$ &0.136\\
 7620& B & 0.480 &0.6562&$-2.05$ &0.104\\
10811& B & 0.500 &0.4764&$-1.42$ &0.132\\
14449& B & 0.523 &0.5841&$-1.70$ &0.145\\
19037& B & 0.505 &0.4113&$-1.26$ &0.144\\
22917& B & 0.485 &0.5647&$-1.34$ &0.092\\
23502& B & 0.525 &0.4722&$-1.43$ &0.159\\
24089& B & 0.474 &0.5598&$-1.31$ &0.080\\
\tableline
 5167& A & 0.514 &0.6302&$-1.32 $ &0.115\\
 8220& A & 0.562 &0.6768&$-1.03$ &0.140\\
 9660& A & 0.496&0.6218&$-1.35$ &0.100\\
 3054& B & 0.433 &0.5080&$-0.94$ &0.048\\
 3412& B & 0.522 &0.5302&$-1.62$ &0.158\\
 4540& B & 0.472 &0.5689&$-1.40$ &0.091\\
\tableline
\end{tabular}
\end{center}
\normalsize
\label{t:tab2bis}
\end{table}

\noindent
\begin{table}[ht]
\vskip 4 cm
\begin{center}
\caption{Instability strip edges defined by RR Lyrae variables in our LMC 
fields\tablenotemark{a}}
\vspace{0.5cm}
\scriptsize
\begin{tabular}{ccccccccccc}
\tableline
\tableline
Id.&Field&$<{\rm B}>-<{\rm V}>$&$<{\rm B}>-<{\rm V}>$&
&&Id.&Field&$<{\rm B}>-<{\rm V}>$&$<{\rm B}>-<{\rm V}>$&$<{\rm B}>-<{\rm V}>$\\ 
&      &int.&mag.&&&&&int. &mag.  & adopted\\
\tableline
\multicolumn{4}{c}{First Overtone Blue Edge}&&&
\multicolumn{5}{c}{Fundamental Red Edge}\\
 2517&   B &   0.230 &   0.237 &  &  &  8094&   A &   0.509 &  0.517 & 0.525\tablenotemark{b}\\
 2623&   A &   0.233 &   0.240 &  &  & 28293&   A &   0.502 &  0.508 & 0.514\tablenotemark{b}\\ 
 2223&   A &   0.249 &   0.258 &  &  &  7468&   A &   0.481 &  0.489 & 0.481\tablenotemark{b}\\
 4008&   B &   0.278 &   0.284 &  &  &  5589&   A &   0.475 &  0.480 & 0.480\tablenotemark{b}\\
 7783&   B &   0.285 &   0.298 &  &  &  5452&   A &   0.473 &  0.473 & 0.473\\ 
\tableline
\end{tabular}
\end{center}
\tablenotetext{a}{Colors of the variables in field A have been made bluer 
by 0.03 mag to put them on the same reddening scale of field B.}
\tablenotetext{b}{Corrected colors derived using Walker (1998) 
0.04 mag dex$^{-1}$ shift of the color of 
the FRE with metal abundance, and assuming [Fe/H]=$-$1.66 for M3
(Zinn \& West, 1984).}
\normalsize
\label{t:tab3}
\end{table}

\noindent
\begin{table}[ht]
\begin{center}
\caption{Summary of our photometric results. Errors in the lower portion of the 
table include internal dispersion of the average, contributions of the uncertainty
of the photometric calibration and of the aperture corrections, and absorption 
contribution due to the uncertainty in the reddening}
\vspace{0.5cm}
\begin{tabular}{lrl}
\tableline
\tableline
~~~~~~~~~~~~~~~Field A&&~~~~~~~~~~~~~~Field B\\
$<V(RR)>$=19.412$\pm$ 0.019&&$<V(RR)>$=19.320$\pm$ 0.023\\
$<B(RR)>$=19.807$\pm$ 0.022&&$<B(RR)>$=19.680$\pm$ 0.024\\
&&\\
$<V(clump)>$=19.304$\pm$ 0.002&&$<V(clump)>$=19.291$\pm$ 0.003\\
$<B(clump)>$=20.215$\pm$ 0.003&&$<B(clump)>$=20.194$\pm$ 0.003\\
$<I(clump)>$=18.319$\pm$ 0.002&&$<I(clump)>$=18.307$\pm$ 0.003\\
&&\\
$E(B-V)$=0.116$\pm$0.017&&$E(B-V)$=0.086$\pm$0.017\\
~~~~~~R$_{V}$=3.10&R$_{B}$=4.10&~~~~~~~~~~~R$_{I}$=1.94\\
&&\\
$<V(RR)>_{0}$=19.052&&$<V(RR)>_{0}$=19.053\\
$<B(RR)>_{0}$=19.331&&$<B(RR)>_{0}$=19.327\\
&&\\
$<V(clump)>_{0}$=18.944&&$<V(clump)>_{0}$=19.024\\
$<B(clump)>_{0}$=19.739&&$<B(clump)>_{0}$=19.841\\
$<V(clump)>_{0}$=18.094&&$<V(clump)>_{0}$=18.140\\
&&\\
\tableline
&&\\
\multicolumn{3}{c}{Average values}\\
\multicolumn{3}{c}{$<V(RR)>_{0}$=19.05$\pm$ 0.06}\\
\multicolumn{3}{c}{$<B(RR)>_{0}$=19.33$\pm$ 0.08}\\
\multicolumn{3}{c}{$<I(clump)>_{0}$=18.12$\pm$ 0.06}\\
&&\\
\tableline
\end{tabular}
\end{center}
\label{t:tab4}
\end{table}

\noindent
\begin{table}[ht]
\vskip 4 cm
\begin{center}
\caption{Comparison with the $<$V(RR)$>_0$ values in the literature
 transformed to a common value of [Fe/H]=$-$1.5}
\vspace{0.5cm}
\scriptsize
\begin{tabular}{lcllcccc}
\tableline
\tableline
$<$V(RR)$>_0$&N$_{\rm stars}$&\multicolumn{4}{c}{Error}&Reddening&Reference\\
&&(1)&~~(2)&(3)&(total)&&\\
\tableline
19.06&108&0.02&~~0.03&0.05&0.06&0.086-0.116&This paper\\
19.04&182&0.04&0.025&0.05&0.07&0.09&Walker (1992a)\\
19.07\tablenotemark{a}&160&0.04&0.025&0.05&0.07&0.07&Walker (1992a)\\
19.16\tablenotemark{b}&680& --   &0.073\tablenotemark{c}&0.06\tablenotemark{c}
&$>$0.09\tablenotemark{c}&0.1&A00--MACHO\\
18.92&71& -- &0.015\tablenotemark{d}&0.08\tablenotemark{d}&$>$0.08&
0.144\tablenotemark{e}&
Udalski et al. (2000)--OGLE-II LMC\_SC21\\
18.93& -- &0.01&0.015\tablenotemark{d}&0.08\tablenotemark{d}&0.08&
0.143\tablenotemark{f}&
Udalski (2000b)--OGLE-II\\
\tableline
\end{tabular}
\end{center}
(1) standard deviation of the average, (2) photometric zero-point, 
(3) absorption contribution
\tablenotetext{a}{Average value without NGC1841} 
\tablenotetext{b}{From A00 $V$=19.45 corrected for a reddening 
of $E(B-V)$=0.1, as in A96, and for the metallicity}
\tablenotetext{c}{These errors are our guess, according to the procedure 
described in Section 6}
\tablenotetext{d}{Sources for the photometric zero-point and 
absorption 
contributions to the total error are Udalski et al. (1998) and Udalski 
(1998a), respectively} 
\tablenotetext{e}{Reddening is the average of the values for 
field LMC\_SC21 in Table 2 of Udalski et al. (1999a). The 18.92 is obtained
subtracting 3.1$\times$0.144=0.446 mag from $<V(RR)>$=19.37 mag quoted in Udalski et 
al. (2000), and withouth metallicity correction since RR Lyrae's in 
OGLE-II field LMC\_SC21 can be expect to share the same average 
metallicity of our field A} 
\tablenotetext{f}{Reddening is average of the values for 
all the LMC fields in Table 2 of Udalski et al. (1999a)}
\normalsize
\label{t:tab5}
\end{table}

\begin{table}[ht]
\caption{Distance moduli to the LMC from Population I and II 
distance indicators}
\begin{center}
\scriptsize
\begin{tabular}{lrl}
\tableline
\tableline
~~~~~~Method&$\mu_{LMC}$&~~~Reference\\
\tableline
\multicolumn{3}{c}{Population I distance indicators}\\
Cepheids: trig. paral.& 18.70$\pm$0.16\tablenotemark{a}&Feast \& Catchpole (1997)\\
Cepheids: MS fitting& 18.55$\pm$0.06\tablenotemark{b}&Laney \& Stobie (1994)\\
Cepheids: B-W& 18.55$\pm$0.10&Gieren et al. (1998), Di Benedetto (1997)\tablenotemark{c}\\
Cepheids: P/L relation& 18.575$\pm$0.2&Groenewegen (2000)\\
Eclipsing Binaries&18.4~$\pm$0.1~&Ribas \& al. (2002)\\
Clump&18.42$\pm$0.07&this paper, see Section 6.2\\
Clump&18.45$\pm$0.07&this paper, see Section 6.2\\
Clump&18.59$\pm$0.09&Romaniello et al. (2000)\\
Clump&18.471$\pm$0.12&Pietrzynsky \& Gieren (2002)\\
Clump&18.54$\pm$0.10&Sarajedini et al. (2002)\\
Miras&18.54$\pm$0.18&Van Leeuwen (1997)\\
Miras&18.64$\pm$0.14&Feast (2000)\\
SN1987a&18.58$\pm$0.05&Panagia (1998)\\
&&\\
\multicolumn{3}{c}{Population II distance indicators}\\
&&\\
GCs: MSF&18.45$\pm$0.09&Gratton et al. (2002a) \& this paper\\
GCs: dyn. mod.&18.45$\pm$0.13&Carretta et al. (2000b) \& this paper\\
HB: trig. paral.&18.44$\pm$0.13&Carretta et al. (2000b) \& this paper\\
RR Lyr: stat. paral.&18.30$\pm$0.14&this paper, see Section 7\\
RR Lyr: BW&18.38$\pm$0.16&this paper, see Section 7\\
RR Lyr: BW revised&18.50$\pm$0.16& Cacciari et al. (2000) \& this paper\\
LMC RRd's&18.55$\pm$0.19&A97 and A00\\
LMC RRd's&18.52$\pm$0.18&Kovacs (2000)  \& this paper\\
WD cool. seq.&18.40$\pm0.15$&Gratton et al. (2002a) \& this paper\\
RRLyr: trig. par&18.45$\pm$0.16&Benedict et al. (2002) \& this paper\\
TRGB&18.55$\pm$0.09&Cioni et al. (2000)\\
TRGB&18.69$\pm$0.26&Romaniello et al. (2000)\\
\tableline
\end{tabular}
\end{center}
\tablenotetext{a}{Error bar has been revised according 
to Carretta et al. (2000b).}
\tablenotetext{b}{Laney \& Stobie (1994) original value corrected to account for
Hipparcos new distance modulus for the Hyades (see Carretta et al. 2000b).}
\tablenotetext{c}{Average of Gieren et al. (1998) value corrected according 
to Carretta et al (2000b), and of Di Benedetto (1997) estimate.}
\normalsize
\label{t:tab6}
\end{table}

\end{document}